\documentclass[aps,prd,amsmath,amssymb,nofootinbib,notitlepage, 12pt]{revtex4}

\usepackage{stmaryrd}
\usepackage{comment}
\usepackage{color}
\usepackage{amsmath}
\usepackage{amssymb}
\usepackage{amsthm}
\usepackage{mathrsfs}
\usepackage{graphicx}
\usepackage{marvosym}
\usepackage{amsmath,bm}
\usepackage{array}
\usepackage{latexsym}
\usepackage{enumerate}
\usepackage{slashed}
\usepackage{hyperref}
\usepackage{color}
\usepackage{setspace}
\usepackage[normalem]{ulem}

\begin{document}

\title{Quantum kinetic theory for spin transport of quarks with background chromo-electromagnetic fields}
\author{Di-Lun Yang}
\affiliation{Institute of Physics, Academia Sinica, Taipei, 11529, Taiwan}%
\begin{abstract}
We derive the quantum kinetic equations for massive and massless quarks coupled with the background chromo-electromagnetic fields from the Wigner-function approach with the $\hbar$ expansion and effective power-counting scheme. For each case, one obtains coupled color-singlet and color-octet kinetic equations, which also involve the scalar and axial-vector components for the charge and spin transport. These kinetic equations delineate entangled evolution of the corresponding distribution functions decomposed in color space. At weak coupling, we derive the close form of the color-singlet kinetic equations for spin transport, which incorporates the diffusion term and the source term that triggers dynamical spin polarization led by correlation functions of color fields. Also, the non-dynamical source term is found in the axial Wigner function. The induced spin polarization and axial charge currents by these source terms are discussed under physical assumptions for color-field correlators in near-equilibrium quark gluon plasmas.
In the constant-field limit, we further obtain non-vanishing axial Ward identities, from which we extract the pseudo-scalar condensate for massive quarks at finite temperature.     
\end{abstract}
\maketitle
\tableofcontents
\section{Introduction}
In recent years, there have been intensive studies on the developments and applications of quantum kinetic theory (QKT) for spin transport of relativistic fermions. In order to study non-equilibrium transport phenomena beyond e.g. the renown chiral magnetic effect (CME) \cite{Vilenkin:1980fu,Nielsen:1983rb,Alekseev:1998ds,Kharzeev:2004ey,Kharzeev:2007jp,Fukushima:2008xe} and chiral vortical effect (CVE) \cite{Vilenkin:1979ui,Erdmenger:2008rm,Banerjee:2008th,Son:2009tf,Landsteiner:2011cp} in chiral matter composed of massless Weyl fermions, the chiral kinetic theory (CKT) as a QKT that captures the chiral anomaly \cite{Adler:1969gk,Bell:1969ts} and spin-orbit interaction could be constructed by introducing the Berry phase as quantum corrections from the adiabatic approximation \cite{Son:2012wh,Stephanov:2012ki,Chen:2013iga,Chen:2014cla,Chen:2015gta}. Alternatively, by employing the Wigner-function approach based on quantum field theories with the $\hbar$ expansion \cite{Gao:2012ix,Chen:2012ca,Son:2012zy}, the covariant CKT with systematic inclusions of both background electromagnetic fields and collisions involving quantum corrections has been later obtained \cite{Hidaka:2016yjf,Hidaka:2017auj}. There is a considerable number of related and follow-up work \cite{Manuel:2013zaa,Manuel:2014dza,Hidaka:2018ekt,Mueller:2017arw,Mueller:2017lzw,Huang:2018wdl,Yang:2018lew,Liu:2018xip,Lin:2019ytz,Lin:2019fqo,Shi:2020htn,Fauth:2021nwe,Chen:2021azy,Luo:2021uog,Fang:2022ttm}. In phenomenology, the CKT was broadly applied to relativistic heavy ion collisions \cite{Sun:2017xhx,Sun:2018idn,Liu:2019krs}, Weyl semimetals \cite{Gorbar:2016ygi,Gorbar:2016sey,Gorbar:2021ebc}, and neutrino transport in core-collapse supernovae \cite{Yamamoto:2020zrs,Yamamoto:2021hjs}.   

On the other hand, due to the observations of spin polarization of Lambda hyperons and spin alignments of vector mesons in heavy-ion experiments at RHIC and LHC \cite{STAR:2017ckg,STAR:2019erd,ALICE:2019aid,Singha:2020qns} motivated by Refs.~\cite{Liang:2004ph,Liang:2004xn,Becattini2013a} in theory, further studies on the generalization of CKT to the QKT for massive fermions has been carried out in order to understand the dynamical spin polarization of strange quarks traveling through the quark gluon plasmas (QGP). In light of the Wigner-function approaches, such a collisionless QKT to track entangled spin and charge transport has been derived in Refs.~\cite{Gao:2019znl,Weickgenannt:2019dks,Hattori:2019ahi,Wang:2019moi}. Later, the systematic inclusion of collisions with quantum corrections was achieved by using the Kadanoff-Baym approach \cite{Yang:2020hri,Sheng:2021kfc} and the extended phase space with non-local collisions \cite{Weickgenannt:2020aaf,Weickgenannt:2021cuo}. See also Refs.~\cite{Florkowski:2018ahw,Zhang:2019xya,Kapusta:2019sad,Liu:2020flb,Li:2019qkf,Wang:2020pej,Wang:2021qnt} for related studies of QKT and spin transport. In particular, from detailed balance of effective models, the vanishing collision term in QKT yields the Wigner function with vorticity correction that gives rise to spin polarization in global equilibrium \cite{Weickgenannt:2020aaf,Wang:2020pej}. Such a result matches the one previously found from different methods \cite{Becattini2013a,Fang:2016vpj} that provides the modified Cooper-Frye formula dictated by thermal vorticity, which successfully describes the global spin polarization observed by STAR collaboration after implementing numerical simulations \cite{Karpenko:2016jyx,Li:2017slc,Xie:2017upb,Wei:2018zfb,Ryu:2021lnx}. In fact, the vanishing collision term of the CKT with Coulomb scattering yields the Wigner functions in local equilibrium, which incorporates extra corrections such as shear and chemical-potential gradient effects \cite{Hidaka:2017auj}. These corrections for massive fermions were later obtained by using the linear response theory \cite{Liu:2020dxg,Liu:2021uhn} and the statistical-field theory \cite{Becattini:2021suc} (see also Refs.\cite{Yi:2021ryh,Liu:2021nyg}). Especially, the shear correction could lead to a substantial contributions on local spin polarization \cite{Fu:2021pok,Becattini:2021iol,Yi:2021ryh,Florkowski:2021xvy,Yi:2021unq,Sun:2021nsg}, which is crucial to explain the experimental measurements \cite{STAR:2019erd}. However, it is also found that the local polarization is sensitive to the equation of state and freeze-out temperature \cite{Becattini:2021iol,Yi:2021ryh}. Therefore, further studies involving non-equilibrium corrections from collisions are still required.     

Nevertheless, the role of gluons or color degrees of freedom that may potentially affect the spin transport of quarks has been relatively overlooked in previous studies. Even though the thermalized gluons could lead to the spin diffusion through classical collisional effects shown in Refs.~\cite{Li:2019qkf,Yang:2020hri}, the quantum correction as a source term triggering the spin polarization from gluons is currently unknown\footnote{There are studies for the construction of the QKT for polarized photons with possible generalization to weakly-coupled gluons \cite{Huang:2020kik,Hattori:2020gqh}, while the direct application to QGP is not yet feasible.}. On the other hand, in addition to the scattering with on-shell gluons, there exist dynamically generated chromo-electromagnetic fields originating from Weibel-type instabilities in expanding QGP \cite{Mrowczynski:1988dz,Mrowczynski:1993qm,Romatschke:2003ms} that may potentially influence the spin transport of quarks in heavy ion collisions. Such color fields could result in anomalous dissipative transport that may dominate over the collisional effects in weakly coupled quantum chromodynamics (QCD) as proposed in Refs.~\cite{Asakawa:2006tc,Asakawa:2006jn}. See also Refs.~\cite{Mrowczynski:2017kso,Carrington:2020sww} for the analogous effect upon heavy-quark transport. The possible impact upon spin transport of quarks from color fields has been recently found by using a similar analysis \cite{Muller:2021hpe}, which in particular stems from the parity-odd correlators of color fields that could be potentially generated in different phases of heavy ion collisions from distinct mechanisms \cite{Joyce:1997uy,Akamatsu:2013pjd,Mace:2016svc,Tanji:2016dka,Mace:2019cqo}. In this paper, we present the derivation of QKT for massive quarks coupled with background color fields applied in Ref.~\cite{Muller:2021hpe} via the Wigner-function approach and effective power counting. More general results and details for the derivation of source terms in color-singlet Wigner functions and kinetic equations responsible for spin polarization are shown. Furthermore, we investigate the axial Ward identity from the axial charge current led by parity-odd color-field correlators in the constant-field limit. In addition, we perform a similar analysis for massless quarks by utilizing the CKT with background color fields constructed in Ref.~\cite{Luo:2021uog}. 

Since the application of the derived QKT to QGP relies on several assumptions, here we briefly summarize the validity and critical approximations of our approach. Similar to the generic QKT with $\hbar$ expansion, it is required the energy of a quasi-particle is much larger than the gradient scale, $\epsilon_{\bm p}\gg \mathcal{O}(\partial)$. For the QKT of massive quarks derived in Sec.~\ref{sec:sol_and_KE}, it is further subject to $m\gg \mathcal{O}(\partial)$ with $m$ being the quark mass due to a technical reason. Our approach is applicable to weakly coupled QCD and the collision term is assumed to be suppressed by sufficiently strong background color fields. Moreover, the other sources for spin polarization pertinent to the fluid properties such as thermal vorticity are assumed to be relatively small and thus neglected. In order to further solve for dynamical spin polarization from kinetic equations, we further assume the correlation function of color fields takes a Gaussian form depending on only the time difference in light of space-time translational invariance and propose a hierarchy for different color-field correlators according to the screening of chromo-electric fields.     
 
This article is organized as follows: In Sec.~\ref{sec:KB_eq}, we derive the master equations from free-streaming Kadanoff-Baym equation for massive quarks under background color fields. In Sec.~\ref{sec:sol_and_KE}, we solve for the perturbative solution and kinetic equations up to $\mathcal{O}(\hbar)$ with effective power counting and further perform the color decomposition to separate the color-singlet and color-octet components. In Sec.~\ref{sec:diffusion_source}, the spin diffusion and source terms for the color-singlet kinetic equations and associated Wigner functions are obtained in weak coupling. In Sec.~\ref{sec:spin_pol_aixal_Ward}, the spin polarization, axial charge current, and axial Ward identity are investigated based on postulated color-field correlators. In Sec.~\ref{sec:massless_fermions}, similar analyses for massless quarks are shown. Finally, in Sec.~\ref{sec:summary}, we make the summary and discussions.

Throughout this paper we use 
the mostly minus signature of the Minkowski metric $\eta^{\mu\nu} = {\rm diag} (1, -1,-1,-1)  $ 
and the completely antisymmetric tensor $ \epsilon^{\mu\nu\rho\lambda} $ with $ \epsilon^{0123} = 1 $. 
We use the notations $A^{(\mu}B^{\nu)}\equiv A^{\mu}B^{\nu}+A^{\nu}B^{\mu}$ 
and $A^{[\mu}B^{\nu]}\equiv A^{\mu}B^{\nu}-A^{\nu}B^{\mu}$. We also define $\tilde{F}^{\mu\nu}\equiv\epsilon^{\mu\nu\alpha\beta}F_{\alpha\beta}/2$.

\section{Master equations from the Kadanoff-Baym equation}\label{sec:KB_eq}
To track the dynamical spin polarization for quarks in heavy ion collisions with non-Abelian color fields in quark gluon plasmas (QGP) or even pre-equilibrium phases, we generalize the so-called axial kinetic theory (AKT) constructed in Refs.~\cite{Hattori:2019ahi,Yang:2020hri}, as a quantum kinetic theory (QKT) to delineate the intertwined dynamics between charge and spin evolution of relativistic fermions, to the case with background color fields in QCD. The AKT incorporates a scalar kinetic equation (SKE) and an axial-vector kinetic equation (AKE) obtained from the Wigner-function approach with the $\hbar$ expansion generally equivalent to a gradient expansion capturing quantum corrections. In addition, we apply a slightly different approach based on Ref.~\cite{Elze:1986qd} by deriving the perturbative solution for Wigner operators and the corresponding kinetic equations in operator form first, while taking ensemble average to obtain the Wigner functions and the quantum-averaged kinetic equations in the end of calculations. 

To consider a more rigorous generalization for the AKT from quantum electrodynamics (QED) to QCD, we have to introduce extended Wigner functions incorporating color degrees of freedom. According to Ref.~\cite{Elze:1986qd}, the gauge-covariant Wigner operator of quarks is defined as
\begin{eqnarray}\label{eq:WT}
\grave{S}^{<}(p,X)=\int\frac{d^4Y}{(2\pi)^4}e^{\frac{ip\cdot Y}{\hbar}}S^{<}(X,Y),
\end{eqnarray}  
where 
\begin{eqnarray}\label{eq:Sless}
S^{<}(X,Y)=\bar{\psi}\left(X-\frac{Y}{2}\right)U\left(-\frac{Y}{2},X\right)\otimes U\left(X,X+\frac{Y}{2}\right)\psi\left(X+\frac{Y}{2}\right)
\end{eqnarray}
as a two-point operator for Dirac fermions. The quantum expectation value of the Wigner operator, $\langle\grave{S}^{<}(p,X)\rangle$, by taking the ensemble average corresponds to the Wigner function. Here $U(x_2,x_1)$ denotes the gauge link with the path integration along a straight line between two end points, $x_2$ and $x_1$, and the tensor product $\otimes$ implies that $S^{<}(X,Y)$ is a matrix in spinor and color spaces. Also, $p_{\mu}$ represents the kinetic momentum. We will focus on the lesser propagators of quarks throughout this paper and hence add the superscript $^<$ for the Wigner operator. On the other hand, the dynamics of $\grave{S}^{<}$ is governed by the Kadanoff-Baym equation (See Ref.~\cite{Hidaka:2022dmn} for a review),  
\begin{eqnarray}\label{KB_collisions}
\bigg(\gamma^{\mu}\hat{\Pi}_{\mu}-m+\frac{i\hbar}{2}\gamma^{\mu}\hat{\nabla}_{\mu}\bigg)\grave{S}^{<}=\frac{i\hbar}{2}\Big(\Sigma^{<}\star S^>-\Sigma^>\star S^<\Big),
\end{eqnarray}
where $\Sigma^{<}$ and $\Sigma^{>}$ represent the lesser and greater self energies in operator form, which characterize the collisional effects. The notation $\star$ denotes the Moyal product, whereas its explicit definition is unimportant here. The explicit expressions of the operators, $\hat{\Pi}_{\mu}$ and $\hat{\nabla}_{\mu}$, will be shown later. In the absence of background fields, they simply reduce to $\hat{\Pi}_{\mu}=p_{\mu}$ and $\hat{\nabla}_{\mu}=\partial_{\mu}$. Here we also neglect the real part of the retarded Wigner operator and of the retarded self energy. The former is dropped for the quasi-particle approximation, while the latter is omitted for simplicity\footnote{The real part of the retarded self energy leads to the modification on the dispersion relation such as the thermal mass. It may further yield a potential term in the kinetic equation when it depends on $X$. See Ref.~\cite{Hidaka:2022dmn} for detailed discussions. How such a term could possibly affect the spin transport requires further studies, which is beyond the scope of current work. } 
At weak coupling, the collision term is of $\mathcal{O}(g^4\ln g)$ (e.g. \cite{Li:2019qkf,Yang:2020hri}), which could be suppressed by sufficiently strong background color fields, superficially of $\mathcal{O}(g)$ and more precisely of $\mathcal{O}(g^2)$ as will be shown in this work, giving rise to the Lorentz force and anomalous force from the $\hbar$ correction. We hence focus on the collisionless case in the present work.      
Now, the free-streaming Kadanoff-Baym equation in QCD, dictating the dynamical evolution of Wigner operators, takes the form \cite{Elze:1986qd}
\begin{eqnarray}
&&\bigg(\gamma^{\mu}p_{\mu}-m+\frac{i\hbar}{2}\gamma^{\mu}D_{\mu}\bigg)\grave{S}^{<}
\\\nonumber
&&=-\frac{i\hbar}{2}\partial_{p}^{\nu}\bigg(\int^1_0ds\Big(1-\frac{s}{2}\Big)\big[e^{-\frac{i\hbar}{2}(1-s)\partial_{p}\cdot D}F_{\nu\mu}\big]\grave{S}^{<}
+\grave{S}^{<}\int^1_0ds\frac{(1-s)}{2}\big[e^{\frac{i\hbar}{2}s\partial_{p}\cdot D}F_{\nu\mu}\big]\bigg)
\\\nonumber
&&=-\frac{i\hbar}{2}\partial_{p}^{\nu}\gamma^{\mu}\bigg(F_{\nu\mu}\grave{S}^{<}
-\frac{1}{4}[F_{\nu\mu},\grave{S}^{<}]_{\rm c}-\frac{i\hbar}{6}\big(\partial_{p}\cdot DF_{\nu\mu}\big)\grave{S}^{<}-\frac{i\hbar}{24}[(\partial_{p}\cdot DF_{\nu\mu}),\grave{S}^{<}]_{\rm c}\bigg)+\mathcal{O}(\hbar^3),
\end{eqnarray}
which can be rearranged as
\begin{eqnarray}
\bigg(\gamma^{\mu}\hat{\Pi}_{\mu}-m+\frac{i\hbar}{2}\gamma^{\mu}\hat{\nabla}_{\mu}\bigg)\grave{S}^{<}=0,
\end{eqnarray}
where 
\begin{eqnarray}\nonumber
\hat{\Pi}_{\mu}\grave{S}^{<}&=&p_{\mu}\grave{S}^{<}+\frac{i\hbar}{8}[F_{\nu\mu},\partial_{p}^{\nu}\grave{S}^{<}]_{\rm c}+\frac{\hbar^2}{12}\big(\partial_{p}\cdot DF_{\nu\mu}\big)\partial_{p}^{\nu}\grave{S}^{<}+\mathcal{O}(\hbar^3),
\\
\hat{\nabla}_{\mu}\grave{S}^{<}&=&D_{\mu}\grave{S}^{<}+\frac{1}{2}\{F_{\nu\mu},\partial_{p}^{\nu}\grave{S}^{<}\}_{\rm c}-\frac{i\hbar}{24}[(\partial_{p}\cdot DF_{\nu\mu}),\partial_p^{\nu}\grave{S}^{<}]_{\rm c}+\mathcal{O}(\hbar^2).
\end{eqnarray}
Here $D_{\mu}O=\partial_{\mu}O+i[A_{\mu},O]_{\rm c}$, $A_{\mu}=t^aA_{\mu}^a$, $F_{\nu\mu}=t^aF^{a}_{\nu\mu}$, and $t^a=\lambda^a/2$ with $\lambda^a$ the Gell-Mann matrices. Also, $\{\,,\,\}_{\rm c}$ and $[\,,\,]_{\rm c}$ denote the anti-commutation and commutation relations in color space. The collision term is of the higher order at weak coupling and hence omitted.

We then apply the decomposition based on the Clifford algebra \cite{Vasak:1987um} in spinor space,
\begin{eqnarray}
	\grave{S}^<=\mathcal{S}+ i\mathcal{P}\gamma^5+ \mathcal{V}^{\mu}\gamma_\mu+\mathcal{A}^{\mu}\gamma^5\gamma_{\mu}+ \frac{\mathcal{S}^{\mu\nu}}{2}\Sigma_{\mu\nu},
\end{eqnarray}
where $\Sigma_{\mu\nu}=i[\gamma_{\mu},\gamma_{\nu}]/2$ and $\gamma^5=i\gamma^0\gamma^1\gamma^2\gamma^3$. In particular, the vector and axial-vector components, $\mathcal{V}^{\mu}(p,X)$ and $\mathcal{A}^{\mu}(p,X)$, directly contributes to the vector charge current and the axial charge current (also spin polarization), receptively. We hence focus on the derivation of these two components. Inserting $\grave{S}^<$ with the decomposition above into the Kadanoff-Baym equation, one should derive 10 master equations. We can further write $\mathcal{S}$, $\mathcal{P}$, and $\mathcal{S}_{\mu\nu}$ in terms of $\mathcal{V}_{\mu}$ and $\mathcal{A}_{\mu}$ via three of the master equations,
\begin{align}
	\label{meq_3}
	m\mathcal{S}&=\hat{\Pi}\cdot\mathcal{V},
	\\\label{meq_4}
	m\mathcal{P}&=- \frac{\hbar}{2}\hat{\nabla}_{\mu}\mathcal{A}^{\mu},
	\\\label{meq_5}
	m\mathcal{S}_{\mu\nu}&=-\epsilon_{\mu\nu\rho\sigma} \hat{\Pi}^{\rho}\mathcal{A}^{\sigma}
	+\frac{\hbar}{2}\hat{\nabla}_{[\mu}\mathcal{V}_{\nu]}
	.
\end{align}
We accordingly obtain the rest of master equations as
\begin{eqnarray}\label{rrr1}
	&&\hat{\nabla}\cdot\mathcal{V}=0,
	\\\label{rrr2}
	&&(\hat{\Pi}_{\mu}\hat{\Pi}\cdot\mathcal{V}-m^2\mathcal{V}_{\mu})
	=-\frac{\hbar}{2}\epsilon_{\nu\mu\rho\sigma}\hat{\nabla}^{\nu}\hat{\Pi}^{\rho}\mathcal{A}^{\sigma}+\frac{\hbar^2}{4}\hat{\nabla}^{\nu}\hat{\nabla}_{[\nu}\mathcal{V}_{\mu]},
	\\\label{rrr3}
	&&\hat{\Pi}\cdot\mathcal{A}=0,
	\\\label{rrr4}
	&&\hat{\Pi}_{\nu}\mathcal{V}_{\mu}-\hat{\Pi}_{\mu}\mathcal{V}_{\nu}
	=\frac{\hbar}{2}\epsilon_{\mu\nu\rho\sigma}\hat{\nabla}^{\rho}\mathcal{A}^{\sigma},
	\\\label{rrr5}
	&&(\hat{\Pi}^2-m^2)\mathcal{A}^{\mu}-\hat{\Pi}_{\sigma}\hat{\Pi}^{\mu}\mathcal{A}^{\sigma}
	=\frac{\hbar}{2}\epsilon^{\mu\nu\rho\sigma}\hat{\Pi}_{\sigma}
	\hat{\nabla}_{\nu}\mathcal{V}_{\rho}-\frac{\hbar^2}{4}\hat{\nabla}^{\mu}\hat{\nabla}\cdot\mathcal{A}
	,
	\\\label{rrr6}
	&&\hbar\hat{\nabla}_{\sigma}\hat{\Pi}^{\sigma}\mathcal{A}^{\mu}
	-\hbar \hat{\nabla}_{\sigma}\hat{\Pi}^{\mu}\mathcal{A}^{\sigma}+\hbar\hat{\Pi}^{\mu}\hat{\nabla}_{\sigma}\mathcal{A}^{\sigma}
	=\frac{\hbar^2}{4}\epsilon^{\mu\nu\rho\sigma}\hat{\nabla}_{\sigma}\hat{\nabla}_{[\nu}\mathcal{V}_{\rho]}
	,
	\\\label{rrr7}
	&&\hbar\big(\hat{\Pi}\cdot\hat{\nabla}\mathcal{V}_{\mu}
	+\hat{\nabla}_{\mu}\hat{\Pi}^{\nu}\mathcal{V}_{\nu}-\hat{\Pi}^{\nu}\hat{\nabla}_{\mu}\mathcal{V}_{\nu}\big)
	=2\epsilon_{\nu\mu\rho\sigma}(\hat{\Pi}^{\nu}\hat{\Pi}^{\rho})\mathcal{A}^{\sigma}.
\end{eqnarray}
See e.g. Refs.~\cite{Hattori:2019ahi,Yang:2020hri} for an analogous derivation in QED. However, for further simplification in QCD, we may adopt the power counting in Ref.~\cite{Yang:2020hri} such that $\mathcal{V}^{\mu}\sim \mathcal{O}(\hbar^0)$ and $\mathcal{A}^{\mu}\sim \mathcal{O}(\hbar)$ due to the quantum nature of $\mathcal{A}^{\mu}$ as the spin-current density in phase space and retain only the leading-order contribution in the $\hbar$ expansion. The master equations then reduce to
\begin{eqnarray}\label{rrr1_sim}
	&&\tilde{\Delta}\cdot\mathcal{V}=0,
	\\\label{rrr2_sim}
	&&p_{\mu}p\cdot\mathcal{V}-m^2\mathcal{V}_{\mu}
	=0,
	\\\label{rrr3_sim}
	&&p_{[\nu}\mathcal{V}_{\mu]}=0,
	\\\label{rrr4_sim}
	&&p\cdot\mathcal{A}=0,
	\\\label{rrr5_sim}
	&&(p^2-m^2)\mathcal{A}^{\mu}
	=\frac{\hbar}{2}\epsilon^{\mu\nu\rho\sigma}p_{\sigma}
	\tilde{\Delta}_{\nu}\mathcal{V}_{\rho}
	,
	\\\label{rrr6_sim}
	&&p\cdot\tilde{\Delta}\mathcal{A}^{\mu}+\frac{1}{2}\{F^{\nu\mu},\mathcal{A}_{\nu}\}_{\rm c}
	=\frac{\hbar}{2}\epsilon^{\mu\nu\rho\sigma}\tilde{\Delta}_{\sigma}\tilde{\Delta}_{\nu}\mathcal{V}_{\rho},
\end{eqnarray}
where $\tilde{\Delta}_{\mu}O=D_{\mu}O+\{F_{\nu\mu}\partial_{p}^{\nu},O\}_{\rm c}/2$ and we have dropped Eq.~(\ref{rrr7}) as a redundant equation that can be obtained from the combination of other master equations. Eqs.~(\ref{rrr1_sim})-(\ref{rrr6_sim}) are same as those in QED by simply replacing the operator $\Delta_{\mu}=\partial_{\mu}+F_{\nu\mu}\partial_{p}^{\nu}$ in QED by $\tilde{\Delta}_{\mu}$ except for further corrections on Eq.~(\ref{rrr6_sim}). By using 
\begin{eqnarray}\nonumber
&&D_{[\sigma}D_{\nu]}O=i[F_{\sigma\nu},O]_{\rm c},
\quad
D_{[\sigma}\{F_{\beta\nu]},\partial_{p}^{\beta},O\}_{\rm c}+\{F_{\beta[\sigma},D_{\nu]}\partial_{p}^{\beta}O\}=\{(D_{[\sigma}F_{\beta\nu]}),\partial^{\beta}_{p}O\}_{\rm c},
\\
&&\{F_{\alpha[\sigma},\{F_{\beta\nu]},\partial^{\alpha}_{p}\partial^{\beta}_{p}O\}\}_{\rm c}
=[F_{\alpha[\sigma}F_{\beta\nu]},\partial^{\alpha}_{p}\partial^{\beta}_{p}O]_{\rm c},
\end{eqnarray}
and hence
\begin{eqnarray}
\tilde{\Delta}_{[\sigma}\tilde{\Delta}_{\nu]}O=
i[F_{\sigma\nu},O]_{\rm c}+\frac{1}{2}\{(D_{[\sigma}F_{\beta\nu]}),\partial^{\beta}_{p}O\}_{\rm c}+\frac{1}{4}[F_{\alpha[\sigma}F_{\beta\nu]},\partial^{\alpha}_{p}\partial^{\beta}_{p}O]_{\rm c},
\end{eqnarray}
Eq.~(\ref{rrr6_sim}) becomes
\begin{eqnarray}\label{rrr6_sim_explicit}
p\cdot\tilde{\Delta}\mathcal{A}^{\mu}+\frac{1}{2}\{F^{\nu\mu},\mathcal{A}_{\nu}\}_{\rm c}
=\frac{\hbar}{4}\epsilon^{\mu\nu\rho\sigma}\Big(\{(D_{\nu}F_{\beta\rho}),\partial^{\beta}_{p}\mathcal{V}_{\sigma}\}_{\rm c}+i[F_{\nu\rho},\mathcal{V}_{\sigma}]_{\rm c}
+\frac{1}{2}[F_{\alpha\nu}F_{\beta\rho},\partial^{\alpha}_{p}\partial^{\beta}_{p}\mathcal{V}_{\sigma}]_{\rm c}\Big).
\end{eqnarray}
Following Ref.~\cite{Hattori:2019ahi}, we can then apply the master equations to derive a perturbative solution of Wigner operators and corresponding kinetic equations.

\section{Perturbative solutions and effective kinetic equations}\label{sec:sol_and_KE}
By using Eqs.~(\ref{rrr1_sim})-(\ref{rrr6_sim}), we will derive the leading-order solution of $\mathcal{V}^{\mu}$ and $\mathcal{A}^{\mu}$ and the corresponding SKE and AKE. Note that we will keep $\hbar$ as a parameter to specify the quantum corrections while taking the power counting for $\mathcal{V}^{\mu}\sim \mathcal{O}(\hbar^0)$ and $\mathcal{A}^{\mu}\sim \mathcal{O}(\hbar)$. Hereafter all terms in equations will be of the same order of the $\hbar$ expansion unless specified otherwise. 

\subsection{Axial kinetic equation}
From Eqs.~(\ref{rrr2_sim}) and (\ref{rrr3_sim}), the vector component now takes the bookkeeping form\footnote{For brevity, we omit the sign function of energy in front of the dispersion relation, which is required to incorporate the contributions from both fermions and anti-fermions.}  
\begin{eqnarray}
	\mathcal{V}^{\mu}=2\pi \delta(p^2-m^2)p^{\mu}\hat{f}_V
	\label{eq:V-LO}
	,
\end{eqnarray}
and the SKE from Eq.~(\ref{rrr1_sim}) reads
\begin{eqnarray}\label{SKE}
\delta(p^2-m^2)p^{\mu}\Big(D_{\mu}\hat{f}_V+\frac{1}{2}\{F_{\nu\mu},\partial_{p}^{\nu}\hat{f}_V\}_{\rm c}\Big)=0,
\end{eqnarray}
which corresponds to the classical kinetic equation in Ref.~\cite{Elze:1986qd}. Here $\hat{f}_V(p,X)$ denotes the vector-charge operator, form which we have $\langle \hat{f}_V(p,X)\rangle=f_{V}(p,X)$ as the vector charge distribution function in phase space by taking the ensemble average.
In light of Refs.~\cite{Hattori:2019ahi,Yang:2020hri}, from Eqs.~(\ref{rrr4_sim}) and (\ref{rrr5_sim}), the axial charge component becomes
\begin{eqnarray}\label{axial_sol}
	\mathcal{A}^{\mu}=2\pi\Big[\delta(p^2-m^2)\Big(\hat{a}^{\mu}+\hbar S^{\mu\nu}_{m(n)}\tilde{\Delta}_{\nu}\hat{f}_V\Big)
	+\frac{\hbar}{2} p_{\nu}\delta'(p^2-m^2)\{\tilde{F}^{\mu\nu},\hat{f}_V\}_{\rm c}\Big],
\end{eqnarray} 
where $\delta'(x)\equiv \partial\delta(x)/\partial x$ and $S^{\mu\nu}_{m(n)}=\epsilon^{\mu\nu\alpha\beta}p_{\alpha}n_{\beta}/(2(p\cdot n+m))$ represents the spin tensor and contributes to the magnetization-current term with $n^{\mu}$ being a timelike frame vector coming from a choice of the spin basis \footnote{Such a choice does not affect physical observables. See e.g. Refs~\cite{Hidaka:2016yjf,Hattori:2019ahi} for discussions.}. Here $\hat{a}^{\mu}(p,X)$ denotes the (effective) spin operator in phase space with the constraint $p\cdot \hat{a}=0$ when the quark is on-shell \footnote{Note that the number-density operator becomes a $2\times 2$ matrix in spin space, where the component proportional to an identity matrix corresponds to $\hat{f}_V$ and the other independent components can be written as $\sigma\cdot \hat{S}$. Here $\sigma^{\mu}$ is the Pauli matrix and $\hat{S}^{\mu}$ is related to $\hat{a}^{\mu}$. We could combine $p^{\mu}$ and $\hat{S}^{\mu}$ to form $\hat{a}^{\mu}$. See Appendix C of Ref.~\cite{Hattori:2019ahi} for the details of construction.}. Similarly, we have $\langle\hat{a}^{\mu}(p,X)\rangle=\tilde{a}^{\mu}(p,X)$ as the (effective) spin four vector. Note that the magnetization-current term is actually obtained by generalization from the case without background fields as in the derivation for QED \cite{Hattori:2019ahi}. Accordingly, one may insert Eq.~(\ref{axial_sol}) into Eq.~(\ref{rrr6_sim_explicit}) to derive the AKE with a general frame vector $n^{\mu}=n^{\mu}(X)$, whereas the full kinetic equation is rather complicated.

Nonetheless, to simplify the problem, we may instead work in the rest frame of massive quarks by choosing $n^{\mu}=n^{\mu}_{r}(p)=p^{\mu}/m$ \cite{Weickgenannt:2019dks} albeit the validity for $m\gg \mathcal{O}(\partial)$. In the rest frame, $S^{\mu\nu}_{m(n_r)}=0$ and hence the magnetization-current term in Eq.~(\ref{axial_sol}) vanishes. Here we compute the $\hbar$ contributions on left-hand side of Eq.~(\ref{rrr6_sim_explicit}), 
\begin{eqnarray}\nonumber
&&\frac{1}{2}p\cdot\tilde{\Delta}\Big(p_{\nu}\delta'(p^2-m^2)\{\tilde{F}^{\mu\nu},\hat{f}_V\}_{\rm c}\Big)
+\frac{\delta'(p^2-m^2)}{2}\{F^{\nu\mu},\{F_{\nu\rho}p^{\rho}, \hat{f}_V\}\}_{\rm c}
\\\nonumber
&&=\frac{\delta'(p^2-m^2)}{2}\Big(p_{\nu}\{(p\cdot D\tilde{F}^{\mu\nu}),\hat{f}_V\}_{\rm c}+p_{\nu}\{\tilde{F}^{\mu\nu},p\cdot\tilde{\Delta}\hat{f}_V\}
+\frac{p^{\rho}p^{\nu}}{2}[[F_{\sigma\rho},\tilde{F}_{\mu\nu}],\partial_{p\sigma}\hat{f}_V]_{\rm c}
\\
&&\quad +\epsilon^{\mu\rho\alpha\beta}p_{\rho}\{F_{\alpha}^{\,\,\nu},\{F_{\beta\nu},\hat{f}_V\}\}_{\rm c}
\Big),
\end{eqnarray}
where we employed 
\begin{eqnarray}
\frac{1}{2}\{F_{\nu\rho},\{\tilde{F}^{\mu\nu},\hat{f}_V\}\}_{\rm c}=\frac{1}{2}\{F^{\nu\mu},\{\tilde{F}_{\rho\nu},\hat{f}_V\}\}_{\rm c}+\epsilon^{\mu\rho\alpha\beta}\{F_{\alpha}^{\,\,\nu},\{F_{\beta\nu},\hat{f}_V\}\}_{\rm c}
\end{eqnarray}
by using the Schouten identity. For the right-hand side, one could show
\begin{eqnarray}
&&\frac{p_{\sigma}p^{\beta}}{2}\delta'(p^2-m^2)\epsilon^{\mu\nu\rho\sigma}\{(D_{\nu}F_{\beta\rho}),\hat{f}_V\}_{\rm c}
\\\nonumber
&&=\frac{1}{2}\delta'(p^2-m^2)\Big(p_{\nu}\{(p\cdot D)\tilde{F}^{\mu\nu},\hat{f}_V\}_{\rm c}+p_{\sigma}p^{\mu}\{D_{\nu}\tilde{F}^{\sigma\nu},\hat{f}_V\}_{\rm c}\Big)
-\frac{p^2}{2}\delta'(p^2-m^2)\{D_{\nu}\tilde{F}^{\mu\nu},\hat{f}_V\}_{\rm c}\Big).
\end{eqnarray}
We then derive a free-streaming effective AKE for massive quarks coupled with background color fields, 
\begin{eqnarray}\label{AKE_nr_2}\nonumber
	0&=&\delta(p^2-m^2)
	\Big(p\cdot\tilde{\Delta}\hat{a}^{\mu}+\frac{1}{2}\{F^{\nu\mu},\hat{a}_{\nu}\}_{\rm c}
	-\frac{i\hbar}{2}[\tilde{F}^{\mu\nu}p_{\nu},\hat{f}_V]_{\rm c}
	-\frac{\hbar\epsilon^{\mu\nu\rho\sigma}p_{\sigma}}{8}[F_{\alpha\nu}F_{\beta\rho},\partial^{\alpha}_{p}\partial^{\beta}_{p}\hat{f}_V]_{\rm c}
	\\\nonumber
	&&	-\frac{\hbar}{4}\epsilon^{\mu\nu\rho\sigma}p_{\rho}\{(D_{\sigma}F_{\beta\nu})
	,\partial_{p}^{\beta}\hat{f}_{V}\}_{\rm c}\Big)
	+\frac{\hbar}{2}\delta'(p^2-m^2)\Big(p_{\nu}\{\tilde{F}^{\mu\nu},p\cdot\tilde{\Delta} \hat{f}_V\}_{\rm c}
	+\epsilon^{\mu\rho\alpha\beta}p_{\rho}\{F_{\alpha}^{\,\,\nu},\{F_{\beta\nu},\hat{f}_V\}\}_{\rm c}
	\\
	&&+\frac{p^{\rho}p^{\nu}}{2}[[F_{\sigma\rho},\tilde{F}_{\mu\nu}],\partial_{p}^{\sigma}\hat{f}_V]_{\rm c}\Big),
\end{eqnarray}
where we took the Bianchi identity $D_{\mu}\tilde{F}^{\mu\nu}=0$ for non-Abelian gauge fields.

\subsection{Color decomposition}
Now, both the SKE and AKE in Eqs.~(\ref{SKE}) and (\ref{AKE_nr_2}) are matrix equations in color space. Nonetheless, only the color-single components of $\mathcal{V}^{\mu}$ and $\mathcal{A}^{\mu}$ ($\hat{f}_V$ and $\hat{a}^{\mu}$) directly contribute to physical observables. For example, the vector and axial charge currents are defined as
\begin{eqnarray}\label{def_currents}
J^{\mu}_{V}=4\int\frac{d^4p}{(2\pi)^4}{\rm tr_c}(\langle \mathcal{V}^{\mu}\rangle), \quad J^{\mu}_{5}=4\int\frac{d^4p}{(2\pi)^4}{\rm tr_c}(\langle\mathcal{A}^{\mu}\rangle),
\end{eqnarray}
by taking the traces over color space and ensemble averages.
Therefore, we may further adopt the color decomposition for kinetic equations.
We first consider the SKE with color decomposition by taking $\hat{f}_V=\hat{f}^{\rm s}_VI+\hat{f}^a_Vt^a$, where $I$ represents an identity matrix in color space. Using
\begin{eqnarray}
[t^a,t^b]_{\rm c}=if^{abc}t^c,\quad
\{t^a,t^b\}_{\rm c}=2\bar{C}_2\delta^{ab}I+d^{abc}t^c,\quad
t^at^b=\bar{C}_2\delta^{ab}I+(d^{abc}+if^{abc})\frac{t^c}{2},
\end{eqnarray}
with $\bar{C}_2=1/(2N_c)$ and $N_c=3$ being number of colors, 
Eq.~(\ref{SKE}) becomes
\begin{eqnarray}\nonumber
0&=&\delta(p^2-m^2)p^{\mu}\Big(\partial_{\mu}(\hat{f}^{\rm s}_V+\hat{f}^a_Vt^a)-gf^{bca}t^aA^b_{\mu}\hat{f}^c_V+t^aF^a_{\nu\mu}\partial_{p}^{\nu}\hat{f}^{\rm s}_V
\\
&&
+\bar{C}_2F^a_{\nu\mu}\partial_{p}^{\nu}\hat{f}^{a}_V+\frac{d^{bca}}{2}t^aF^b_{\nu\mu}\partial_{p}^{\nu}\hat{f}^{c}_V\Big),
\end{eqnarray}
which yields
\begin{eqnarray}\label{SKE_singlet}
&&\delta(p^2-m^2)\mathcal{K}_{\rm s}[\hat{f}_V]=0,
\\\label{SKE_octet}
&&\delta(p^2-m^2)\mathcal{K}_{\rm o}^a[\hat{f}_V]=0,
\end{eqnarray}
where
\begin{eqnarray}
	\mathcal{K}_{\rm s}[O]&\equiv&p^{\mu}\Big(\partial_{\mu}O^{\rm s}+\bar{C}_2F^a_{\nu\mu}\partial_{p}^{\nu}O^{a}\Big),
	\\
	\mathcal{K}_{\rm o}^a[O]&\equiv&p^{\mu}\Big(\partial_{\mu}O^a-f^{bca}A^b_{\mu}O^c+F^a_{\nu\mu}\partial_{p}^{\nu}O^{\rm s}
	+\frac{d^{bca}}{2}F^b_{\nu\mu}\partial_{p}^{\nu}O^{c}\Big),
\end{eqnarray}
for an arbitrary color object that can be decomposed into $O=IO^{\rm s}+t^aO^a$.
Eq.~(\ref{SKE_singlet}) and (\ref{SKE_octet}) correspond to the Vlasov equations for the color-singlet and color-octet components, respectively.
Although the prefactors of the force terms in Eqs.~(\ref{SKE_singlet}) and (\ref{SKE_octet}) are distinct from those in Refs.~\cite{Heinz:1983nx,Heinz:1984yq} due to different color decompositions for $\hat{f}_V$, the combination of two kinetic equations still leads to equivalent results in physics.  

Analogously, we may carry out the color decomposition for the AKE. For simplicity, we focus on the AKE in the rest frame. In addition, we may apply the off-shell solution of $\hat{f}_V$ such that $p\cdot\tilde{\Delta}\hat{f}_V=0$.\footnote{In practice, one has to solve for $\hat{f}_V$ and $\hat{a}^{\mu}$ from the kinetic equations and input them into Wigner operators to evaluate physical quantities. Introducing the on-shell $\hat{f}_V$ and $\hat{a}^{\mu}$ first in kinetic equations or later in Wigner operators should not affect the final results.} It is more convenient to denote Eq.~(\ref{AKE_nr_2}) in the following form,
\begin{eqnarray}
\delta(p^2-m^2)\Big(\hat{\Pi}^{\mu\nu}\hat{a}_{\nu}+\hbar\hat{\chi}^{\mu}\hat{f}_{V}\Big)
+\delta'(p^2-m^2)\hbar\hat{\Theta}^{\mu}\hat{f}_V=0,
\end{eqnarray}
where 
\begin{eqnarray}\nonumber
\hat{\Pi}^{\mu\nu}\hat{a}_{\nu}&=& p\cdot\tilde{\Delta}\hat{a}^{\mu}+\frac{1}{2}\{F^{\nu\mu},\hat{a}_{\nu}\}_{\rm c},
\\\nonumber
\hat{\chi}^{\mu}\hat{f}_{V}&=& -\frac{1}{4}\Big(2i[\tilde{F}^{\mu\nu}p_{\nu},\hat{f}_V]_{\rm c}
+\frac{\epsilon^{\mu\nu\rho\sigma}p_{\sigma}}{2}[F_{\alpha\nu}F_{\beta\rho},\partial^{\alpha}_{p}\partial^{\beta}_{p}\hat{f}_V]_{\rm c}
+\epsilon^{\mu\nu\rho\sigma}p_{\rho}\{(D_{\sigma}F_{\beta\nu})
,\partial_{p}^{\beta}\hat{f}_{V}\}_{\rm c}\Big),
\\
\hat{\Theta}^{\mu}\hat{f}_V&=&\frac{\epsilon^{\mu\rho\alpha\beta}}{2}p_{\rho}\{F_{\alpha}^{\,\,\nu},\{F_{\beta\nu},\hat{f}_V\}\}_{\rm c}
+\frac{p^{\rho}p^{\nu}}{4}[[F_{\sigma\rho},\tilde{F}_{\mu\nu}],\partial_{p}^{\sigma}\hat{f}_V]_{\rm c}.
\end{eqnarray}
These terms above without and with $\hbar$ correspond to the classical contributions and quantum corrections in the AKE \footnote{More precisely, the quantum correction in our context refers to the gradient correction. The classical part mentioned here should implicitly encode the quantum effect that generates nonzero $\hat{a}^{\mu}$, while such a term could also exist if we assume there is an external source irrelevant to the quantities of $\mathcal{O}(\partial)$.}, although they are of the same order in our power counting. 
Taking $\hat{a}^{\mu}=\hat{a}^{\rm s\mu}I+\hat{a}^{a\mu }t^a$, it is found
\begin{eqnarray}\nonumber
\hat{\Pi}^{\mu\nu}\hat{a}_{\nu}&=&
\bigg[p\cdot\partial\big(\hat{a}^{\rm s\mu}+\hat{a}^{a\mu}t^a\big)-t^ap^{\nu}f^{bca}A^b_{\nu}\hat{a}^{\mu}_c
+\frac{p^{\nu}}{2}\big(2t^aF^a_{\rho\nu}\partial_{p}^{\rho}\hat{a}^{\rm s\mu}
+2\bar{C}_2F^a_{\rho\nu}\partial_{p}^{\rho}\hat{a}^{a\mu}
\\\nonumber
&&+d^{bca}t^aF^b_{\rho\nu}\partial_{p}^{\rho}\hat{a}^{\mu}_{c}\big)\bigg]
\\
&=&\mathcal{K}_{\rm s}[\hat{a}^{\mu}]+t^a\mathcal{K}_{\rm o}^a[\hat{a}^{\mu}].
\end{eqnarray}
For part of the quantum corrections, one finds
\begin{eqnarray}\nonumber
\hat{\chi}^{\mu}\hat{f}_{V}&=& \frac{1}{4}\bigg\{2t^af^{bca}\tilde{F}^{b\mu\nu}p_{\nu}\hat{f}^c_V
+\frac{\epsilon^{\mu\nu\rho\sigma}p_{\sigma}}{4}t^af^{dea}f^{bcd}F^b_{\alpha\nu}F^c_{\beta\rho}\partial^{\alpha}_{p}\partial^{\beta}_{p}\hat{f}^e_V
\\\nonumber
&&-\epsilon^{\mu\nu\rho\sigma}p_{\rho}
\Big[\big((\partial_{\sigma}F^a_{\beta\nu})-f^{bca}A^b_{\sigma}F^c_{\beta\nu}\big)t^a\partial_{p}^{\beta}\hat{f}^{\rm s}_V
+\big((\partial_{\sigma}F^d_{\beta\nu})-f^{bcd}A^b_{\sigma}F^c_{\beta\nu}\big)
\big(2\bar{C}_2\partial_{p}^{\beta}\hat{f}^d_V
\\
&&+d^{dea}t^a\partial_{p}^{\beta}\hat{f}^e_V\big)
\Big]
\bigg\},
\end{eqnarray} 
which can be rearranged as
\begin{eqnarray}
\hat{\chi}^{\mu}f_{V}=\mathcal{Q}^{\mu}_{\rm s}[\hat{f}_V]+t^a\mathcal{Q}^{a\mu}_{\rm o}[\hat{f}_V],
\end{eqnarray}
where
\begin{eqnarray}
\mathcal{Q}^{\mu}_{\rm s}[O]\equiv
-\frac{1}{2}\big[\epsilon^{\mu\nu\rho\sigma}p_{\rho}
\big((\partial_{\sigma}F^a_{\beta\nu})-f^{bca}A^b_{\sigma}F^c_{\beta\nu}\big)
\bar{C}_2\partial_{p}^{\beta}O^a
\big],
\end{eqnarray}
and
\begin{eqnarray}\nonumber
\mathcal{Q}^{a\mu}_{\rm o}[O]&\equiv &\frac{1}{4}\bigg[2f^{bca}\tilde{F}^{b\mu\nu}p_{\nu}O^c
+\frac{\epsilon^{\mu\nu\rho\sigma}p_{\sigma}}{4}f^{dea}f^{bcd}F^b_{\alpha\nu}F^c_{\beta\rho}\partial^{\alpha}_{p}\partial^{\beta}_{p}O^e
\\
&&-\epsilon^{\mu\nu\rho\sigma}p_{\rho}
\bigg(2\big((\partial_{\sigma}F^a_{\beta\nu})-f^{bca}A^b_{\sigma}F^c_{\beta\nu}\big)\partial_{p}^{\beta}O_{\rm s}+\big((\partial_{\sigma}F^d_{\beta\nu})-f^{bcd}A^b_{\sigma}F^c_{\beta\nu}\big)
d^{dea}\partial_{p}^{\beta}O^e
\bigg)\bigg].
\end{eqnarray}
Finally, it is found $\hat{\Theta}^{\mu}\hat{f}_V=t^a\hat{\Theta}^{a\mu}_{\rm o}[\hat{f}_V]$, where
\begin{eqnarray}
\hat{\Theta}^{a\mu}_{\rm o}[\hat{f}_V]=\frac{\epsilon^{\mu\rho\alpha\beta}}{2}p_{\rho}\big(2\bar{C}_2F^{a\nu}_{\alpha}F_{\beta\nu}^b\hat{f}_V^b+d^{abc}d^{dec}F^{b\nu}_{\alpha}F^d_{\beta\nu}\hat{f}_V^e\big)
-\frac{p^{\rho}p^{\nu}}{4}f^{cda}f^{ebc}F^e_{\sigma\rho}F^b_{\mu\nu}\partial_{p}^{\sigma}\hat{f}^d_V.
\end{eqnarray}
Here we have used the symmetric property of $d^{abc}$ and it turns out that only part of the color-octet components of $\hat{\Theta}^{\mu}\hat{f}_V$ remains. Accordingly, the color-singlet and color-octet components of the AKE read
\begin{eqnarray}\label{AKE_singlet}
&&\delta(p^2-m^2)\big(\mathcal{K}_{\rm s}[\hat{a}^{\mu}]+\hbar\mathcal{Q}^{\mu}_{\rm s}[\hat{f}_V]\big)=0,
\\\label{AKE_octet}
&&\delta(p^2-m^2)\big(\mathcal{K}_{\rm o}^a[\hat{a}^{\mu}]+\hbar \mathcal{Q}^{a\mu}_{\rm o}[\hat{f}_V]\big)+\hbar \delta'(p^2-m^2)\hat{\Theta}^{a\mu}_{\rm o}[\hat{f}_V]=0.
\end{eqnarray}

\section{Spin diffusion and source terms}\label{sec:diffusion_source}
Given Eqs.~(\ref{SKE_singlet}), (\ref{SKE_octet}), (\ref{AKE_singlet}), and (\ref{AKE_octet}), we may follow the approach in Refs.~\cite{Asakawa:2006tc,Asakawa:2006jn} to combine the color-singlet and color-octet kinetic equations into a single equation for $\hat{f}^{\rm s}_V$ or for $\hat{a}^{\rm s\mu}$ with effective diffusion terms and quantum corrections up to $\mathcal{O}(\hbar)$ in weak coupling. The same derivation has been presented in Ref.~\cite{Muller:2021hpe}. More details for the derivation are shown below.

Considering QCD at weak coupling and assuming the color-octet distribution functions are subleading such that $O^a\sim\mathcal{O}(g)$, we may approximate
\begin{eqnarray}\label{Ko_approx}
	\mathcal{K}_{\rm o}^a[O]&\approx&p^{\mu}\Big(\partial_{\mu}O^a-f^{bca}A^b_{\mu}O^c+F^a_{\nu\mu}\partial_{p}^{\nu}O^{\rm s}\Big),
	\\\label{Qo_approx}
\mathcal{Q}^{a\mu}_{\rm o}[O]&\approx &-\frac{1}{2}\Big(\epsilon^{\mu\nu\rho\sigma}p_{\rho}
(\partial_{\sigma}F^a_{\beta\nu})\partial_{p}^{\beta}O^{\rm s}
\Big),
\end{eqnarray}
up to $\mathcal{O}(g)$ except for maintaining the $f^{bca}A^b_{\mu}$ term associated with the gauge link. Whereby Eq.~(\ref{SKE_octet}) gives rise to  
\begin{eqnarray}\label{fV_octet_sol}
\hat{f}_{V}^a(p,X)=-i\int d^4k\int\frac{d^4X'}{(2\pi)^4}U^{ab}(X,X')\frac{e^{ik\cdot(X'-X)}}{p\cdot k+i\epsilon}p^{\mu}F^b_{\nu\mu}(X')\partial_{p}^{\nu}\hat{f}^{\rm s}_V(p,X'),
\end{eqnarray}
where 
\begin{eqnarray}
U^{ac}(X,X')={\rm exp}\bigg[P\bigg(\int^{X}_{X'}f^{abc}A_{\mu}^b(s)ds^{\mu}\bigg)\bigg]
\end{eqnarray}
parallel transports the gauge field from $X'$ to $X$ with $P$ being the path ordering. 
Similarly, from $\mathcal{K}_{\rm o}^a[\hat{a}^{\mu}]$ and $\mathcal{Q}^{a\mu}_{\rm o}[\hat{f}_V]$ in Eqs.~(\ref{Ko_approx}) and (\ref{Qo_approx}), Eq.~(\ref{AKE_octet}) results in
\begin{eqnarray}\nonumber\label{a_octet_sol}
\hat{a}^{a\mu}(p,X)&=&-i\int d^4k\int\frac{d^4X'}{(2\pi)^4}U^{ab}(X,X')\frac{e^{ik\cdot(X'-X)}}{p\cdot k+i\epsilon}\bigg[p^{\mu}F^b_{\nu\mu}(X')\partial_{p}^{\nu}\hat{a}^{\rm s\mu}(p,X')
\\
&&-\frac{\hbar}{2}\epsilon^{\mu\nu\rho\sigma}p_{\rho}
(\partial_{\sigma}F^b_{\beta\nu}(X'))\partial_{p}^{\beta}\hat{f}^{\rm s}_V(p,X')
\bigg]
.
\end{eqnarray}

Replacing $\hat{f}_{V}^a$ in Eq.~(\ref{SKE_singlet}) by Eq.~(\ref{fV_octet_sol}) 
with the relation,
\begin{eqnarray}
\frac{1}{p\cdot k+i\epsilon}= -i\pi\delta(p\cdot k)+PV(1/p\cdot k),
\end{eqnarray}
where $PV(x)$ represents the principle value of $x$, one obtains \cite{Asakawa:2006jn}
\begin{align}\label{SKE_signlet_full}
0&=&\delta(p^2-m^2)\bigg(p\cdot\partial \hat{f}_V^{\rm s}(p,X)-\bar{C}_2p^{\mu}F^a_{\nu\mu}(X)\partial_{p}^{\nu}\int^{p}_{k,X'}U^{ab}(X,X')
p^{\beta}F^b_{\alpha\beta}(X')\partial_{p}^{\alpha}\hat{f}^{\rm s}_V(p,X')\bigg),
\end{align}
where 
\begin{eqnarray}
	\int^{p}_{k,X'}\equiv \int d^4k\int\frac{d^4X'}{(2\pi)^4}e^{ik\cdot(X'-X)}\big(\pi\delta(p\cdot k)+iPV(1/p\cdot k)\big).
\end{eqnarray}
When the correlations of field strengths are even functions with respect to $X-X'$, only the imaginary part of $1/(p\cdot k+i\epsilon)$ contributes as considered in Ref.~\cite{Asakawa:2006jn}. However, the real part of $1/(p\cdot k+i\epsilon)$ could also gives a real contribution to $\hat{f}_{V}^a(p,X)$. See appendix A for a detailed discussion.
On the other hand, by inserting Eq.~(\ref{a_octet_sol}) into Eq.~(\ref{AKE_singlet}) and utilizing Eq.~(\ref{fV_octet_sol}), it is found
\begin{eqnarray}\nonumber\label{AKE_signlet_full}
0&=&\delta(p^2-m^2)\bigg\{p\cdot\partial\hat{a}^{\rm s\mu}(p,X)-\bar{C}_2p^{\lambda}F^a_{\kappa\lambda}(X)\partial_{p}^{\kappa}\int^{p}_{k,X'}
U^{ab}(X,X')\bigg[p^{\beta}F^b_{\alpha\beta}(X')\partial_{p}^{\alpha}\hat{a}^{\rm s\mu}(p,X')
\\\nonumber
&&-\frac{\hbar}{2}\epsilon^{\mu\nu\rho\sigma}p_{\rho}
(\partial_{\sigma}F^b_{\beta\nu}(X'))\partial_{p}^{\beta}\hat{f}^{\rm s}_V(p,X')
\bigg]
\\
&&+\frac{\hbar\bar{C}_2}{2}\epsilon^{\mu\nu\rho\sigma}p_{\rho}
(\partial_{\sigma}F^a_{\kappa\nu}(X))
\partial_{p}^{\kappa}
\int^{p}_{k,X'}U^{ab}(X,X')
p^{\beta}F^b_{\alpha\beta}(X')\partial_{p}^{\alpha}\hat{f}^{\rm s}_V(p,X')\bigg\}.
\end{eqnarray}
By taking the ensemble average, we may recast 
Eq.~(\ref{AKE_signlet_full}) into 
\begin{eqnarray}\label{AKE_signlet_simplify}
0=\delta(p^2-m^2)\Big(p\cdot\partial\tilde{a}^{\rm s\mu}(p,X)-\partial_{p}^{\kappa}\mathscr{D}_{\kappa}[\tilde{a}^{s\mu}]
+\hbar\partial_{p}^{\kappa}\big(\mathscr{A}^{\mu}_{\kappa}[f^{\rm s}_{V}]\big)\Big),
\end{eqnarray}
where 
\begin{eqnarray}
	\mathscr{D}_{\kappa}[O]=\bar{C}_2\int^{p}_{k,X'}p^{\lambda}p^{\rho}\langle F^a_{\kappa\lambda}(X)
F^a_{\alpha\rho}(X')\rangle \partial_{p}^{\alpha}O(p,X')
\end{eqnarray}
and
\begin{eqnarray}
	\mathscr{A}^{\mu}_{\kappa}[O]=\frac{\bar{C}_2}{2}
\epsilon^{\mu\nu\rho\sigma}\int^{p}_{k,X'}p^{\lambda}p_{\rho}\Big(\partial_{X'\sigma}\langle F^a_{\kappa\lambda}(X)F^a_{\alpha\nu}(X')\rangle
+\partial_{X\sigma}\langle F^a_{\kappa\nu}(X)F^a_{\alpha\lambda}(X')\rangle\Big)\partial^{\alpha}_{p}O(p,X').
\end{eqnarray}
Here we have dropped the term proportional to $\epsilon^{\mu\nu\rho\sigma}\partial_{\sigma}F^a_{\rho\nu}(X)=\mathcal{O}(g^2)$ and used $\langle \hat{O} \hat{f}^{\rm s}_{V}\rangle=\langle \hat{O}\rangle \langle\hat{f}^{\rm s}_{V}\rangle$ and $\langle \hat{O} \hat{a}^{\rm s\mu}\rangle=\langle \hat{O}\rangle \langle\hat{a}^{\rm s\mu}\rangle$ with $\hat{O}$ an arbitrary operator based on the quasi-particle approximation. Also, we introduce the color-field correlator with the insertion of a gauge link,
\begin{eqnarray}\label{eq:color_field_corr}
\langle F^a_{\kappa\lambda}(X)
F^a_{\alpha\rho}(X')\rangle \equiv \langle F^a_{\kappa\lambda}(X)U^{ab}(X,X')
F^b_{\alpha\rho}(X')\rangle,
\end{eqnarray}
which comes from the ensemble average of the gauge field and color-octet distribution function.
Note that the background color fields in Eq.~(\ref{eq:color_field_corr}) in QGP originates from soft gluons emitted by stochastic sources or induced by the Weibel-type plasma instability, whereas the influence from quantum fluctuations of hard gluons characterized by on-shell gluonic Wigner functions are encoded in $\Sigma^{\lessgtr}$ in the collision term delineating the interaction with the probe strange quark with hard-momentum exchange. Such a scale separation results in the Fokker-Plank equation plus hard scattering kernel in kinetic theory \cite{Ghiglieri:2015ala,Dai:2020rlu}.
Recall that the color-singlet SKE in Eq.~(\ref{SKE_signlet_full}) with the same approximations reads \cite{Asakawa:2006jn}
\begin{eqnarray}\label{SKE_signlet_simplify}
0=\delta(p^2-m^2)\Big(p\cdot\partial f^{\rm s}_{V}(p,X)-\partial_{p}^{\kappa}\mathscr{D}_{\kappa}[f_V^{s}]\Big)=0.
\end{eqnarray}
One finds that both Eqs.~(\ref{AKE_signlet_simplify}) and (\ref{SKE_signlet_simplify}) have similar diffusion terms, $\partial_{p}^{\kappa}\mathscr{D}_{\kappa}[\tilde{a}^{s\mu}]$ and $\partial_{p}^{\kappa}\mathscr{D}_{\kappa}[f_V^{s}]$, while Eq.~(\ref{AKE_signlet_simplify}) further incorporates the quantum correction, $\hbar\partial_{p}^{\kappa}\big(\mathscr{A}^{\mu}_{\kappa}[f^{\rm s}_{V}]\big)$, as a source term for dynamical spin polarization. 

In addition, from Eq.~(\ref{axial_sol}), the color-singlet component of the axial-vector Wigner function, $\langle\mathcal{A}^{\rm s\mu}\rangle={\rm tr_c}(\langle\mathcal{A}^{\mu}\rangle)/N_c$, in the rest frame reads
\begin{eqnarray}\nonumber\label{eq:Asmu_with_totalDp}
	\langle\mathcal{A}^{\rm s\mu}\rangle&=&2\pi\Big[\delta(p^2-m^2)\tilde{a}^{\rm s\mu}
	+\frac{\hbar\bar{C}_2}{2} \big(\partial_{p\nu}\delta(p^2-m^2)\big)\langle\tilde{F}^{a\mu\nu}\hat{f}^{a}_V\rangle\Big]
	\\
	&=&2\pi\bigg[\delta(p^2-m^2)\bigg(\tilde{a}^{\rm s\mu}-\frac{\hbar\bar{C}_2}{2}\langle\tilde{F}^{a\mu\nu}\partial_{p\nu}\hat{f}^{a}_V\rangle\bigg)
	+\frac{\hbar\bar{C}_2}{2} \partial_{p\nu}\big(\delta(p^2-m^2)\langle \tilde{F}^{a\mu\nu}\hat{f}^{a}_V\rangle\big)\bigg]
	.
\end{eqnarray}
In order to obtain the spectrum of spin polarization for on-shell fermions, we have to integrate over $p\cdot \bar{n}$ with $\bar{n}^{\mu}=(1,{\bm 0})$ being a timelike vector to define the particle energy. It turns out that
\begin{eqnarray}
	\int \frac{dp_0}{2\pi}\langle\mathcal{A}^{\rm s\mu}\rangle=\frac{1}{2\epsilon_{\bm p}}\bigg(\tilde{a}^{\rm s\mu}-\frac{\hbar\bar{C}_2}{2}\langle\tilde{F}^{a\mu\nu}\partial_{p\nu}\hat{f}^{a}_V\rangle\bigg)_{p_0=\epsilon_{\bm p}}
	+\frac{\hbar\bar{C}_2}{4}\partial_{p_{\perp}\nu}\left(\frac{\langle\tilde{F}^{a\mu\nu}\hat{f}^{a}_V\rangle}{\epsilon_{\bm p}}\right)_{p_0=\epsilon_{\bm p}},
\end{eqnarray}
where $p_{\perp}^{\mu}\equiv (\eta^{\mu\nu}-\bar{n}^{\mu}\bar{n}^{\nu})p_{\nu}$, $|\bm p|=\sqrt{-p_{\perp}^2}$, and $\epsilon_{\bm p}\equiv\sqrt{|\bm p|^2+m^2}$. Here we only consider the particle with positive energy. Consequently, we could introduce the color singlet of the on-shell axial charge current density in phase space, 
\begin{eqnarray}\label{eq:Asmu_massive}
	\mathcal{A}^{\rm s\mu}(\bm p,X)\equiv \int \frac{dp_0}{2\pi}\langle\mathcal{A}^{\rm s\mu}\rangle
	=\frac{1}{2\epsilon_{\bm p}}\big(\tilde{a}^{\rm s\mu}+\hbar\bar{C}_2\mathcal{A}^{\mu}_{Q}\big)_{p_0=\epsilon_{\bm p}},
\end{eqnarray} 
where $\mathcal{A}^{\mu}_{Q}=\mathcal{A}^{\mu}_{Q1}+\mathcal{A}^{\mu}_{Q2}$ and
\begin{equation}\label{AQmu_origin}
	\mathcal{A}^{\mu}_{Q1}=\bigg[\frac{\partial_{p\kappa}}{2}\int^{p}_{k,X'}p^{\beta}\langle \tilde{F}^{a\mu\kappa}(X)F^a_{\alpha\beta}(X')\rangle\partial_{p}^{\alpha}f^{\rm s}_V(p,X')\bigg]_{p_0=\epsilon_{\bm p}},
\end{equation}
\begin{eqnarray}\nonumber
\mathcal{A}^{\mu}_{Q2}&=&-\frac{\epsilon_{\bm p}}{2}\partial_{p_\perp\kappa}\bigg[\int^{p}_{k,X'}\hat{p}^{\beta}\langle \tilde{F}^{a\mu\kappa}(X)F^a_{\alpha\beta}(X')\rangle\partial_{p}^{\alpha}f^{\rm s}_V(p,X')\bigg]_{p_0=\epsilon_{\bm p}}
\\
&=&\frac{1}{2\epsilon_{\bm p}^2}(p_{\perp\kappa}-\epsilon_{\bm p}^2\partial_{p_{\perp}\kappa})\bigg[\int^{p}_{k,X'}p^{\beta}\langle \tilde{F}^{a\mu\kappa}(X)F^a_{\alpha\beta}(X')\rangle \partial_{p}^{\alpha}f^{\rm s}_V(p,X')\bigg]_{p_0=\epsilon_{\bm p}},
\end{eqnarray}
with $\hat{p}_{\mu}\equiv p_{\mu}/p_0$.

One should solve for $f^{\rm s}_{V}(p,X)$ and $\tilde{a}^{\rm s\mu}$ from Eqs.~(\ref{SKE_signlet_simplify}) and (\ref{AKE_signlet_simplify}), respectively, and input the solutions to Eq.~(\ref{eq:Asmu_massive}) for evaluating the spin polarization spectrum governed by the modified Cooper-Frye formula over the freeze-out hypersurface $\Sigma_{\mu}$ \cite{Becattini2013a,Fang:2016vpj},
\begin{equation}\label{Spin_CooperFrye}
	\mathcal{P}^{\mu}({\bm p})=\frac{\int d\Sigma\cdot p	\hat{\mathcal{A}}^{\rm s\mu}(\bm p,X)}{2m\int d\Sigma\cdot pf_{V}^{\rm s}(\bm p,X)},
\end{equation}
where $\hat{\mathcal{A}}^{\rm s\mu}(\bm p,X)=2\epsilon_{\bm p}\mathcal{A}^{\rm s\mu}(\bm p,X)$ and we have imposed the on-shell condition with positive energy. In Ref.~\cite{Muller:2021hpe}, there is a mistake for the omission of $\mathcal{A}^{\mu}_{Q2}$.

\section{Spin polarization and axial Ward identity}\label{sec:spin_pol_aixal_Ward}
Although the genuine color-field correlators have to be obtained through the real-time simulations in heavy ion collisions, we may propose an approximated form based on the physical arguments as presented in Ref.~\cite{Muller:2021hpe}. Here we provide details for the analysis therein and further discuss the constant-field limit, which is not a realistic situation in heavy ion collisions but an ideal case for theoretical interest. We will focus on the source terms giving rise to spin polarization, while the diffusion term in the SKE has been studied in Ref.~\cite{Asakawa:2006jn} and the one in the AKE takes a similar form. Note that what we study here is a non-equilibrium (more precisely, near-equilibrium) effect on spin polarization in particular due to the involvement of color electric fields. The detailed balance for hard collisions could yield an equilibrium result for spin polarization proportional to thermal vorticity. The anomalous spin polarization from color fields should be regraded as a non-equilibrium correction on top of the equilibrium one, whereas the hydrodynamic gradient terms are assumed to be suppressed compared to the magnitude of color fields in the following analysis. One could in principle include other non-equilibrium corrections from the hard scattering kernel of QKT in e.g. Refs.~\cite{Wang:2021qnt,Fang:2022ttm} with non-QCD effective models. In practice, we focus on high-energy collisions such that the thermal vorticity and other gradient terms are expected to be negligible, which is supported  by the suppressed spin polarization of $\Lambda$ hyperons in LHC.

\subsection{Spin polarization and axial charge currents}
Physically, considering (space-time) translational invariance, we may assume $\langle F^a_{\kappa\lambda}(X)
F^a_{\alpha\rho}(X')\rangle  $ only depends on $X-X'$. More precisely, we may set $\langle F^a_{\kappa\lambda}(X)F^a_{\alpha\rho}(X')\rangle=\langle F^a_{\kappa\lambda}(X')
F^a_{\alpha\rho}(X)\rangle=\langle F^a_{\kappa\lambda}F^a_{\alpha\rho}\rangle\Phi(X-X')$ \footnote{Note that the Bianchi identity $D_{\mu}F^{\mu\nu}=0$ should yields $k^{\mu}\langle \tilde{F}^a_{\mu\nu}F^a_{\alpha\rho}\rangle=k^{\nu}\langle \tilde{F}^a_{\mu\nu}F^a_{\alpha\rho}\rangle=0$ as the constraint under this approximation.}. Introducing new coordinates, $\bar{X}=(X+X')/2$ and $\delta X=X-X'$, one finds $\partial_{X}=\partial_{\bar{X}}/2+\partial_{\delta X}$ and $\partial_{X'}=\partial_{\bar{X}}/2-\partial_{\delta X}$. By employing the relation in Eq.~(\ref{eq:DkintG}), the dynamical source term stemming from quantum corrections in Eq.~(\ref{AKE_signlet_simplify}) becomes
\begin{eqnarray}\nonumber
\partial_{p}^{\kappa}\big(\mathscr{A}^{\mu}_{\kappa}[f^{\rm s}_{V}]\big)&=&-\frac{\bar{C}_2}{2}
\epsilon^{\mu\nu\rho\sigma}\int^{p}_{k,\delta X}
\bigg[\partial_{\delta X\sigma}
\big(p_{\rho}\langle F^{a\lambda}_{\nu}(X)F^a_{\alpha\lambda}(X')\rangle+p^{\lambda}\langle F^{a}_{\rho\lambda}(X)F^a_{\alpha\nu}(X')\rangle\big)\partial_{p}^{\alpha}f^{\rm s}_{V}(p)
\\\nonumber
&&+p^{\lambda}p_{\rho}\partial_{\delta X\sigma}\Big(\langle F^a_{\kappa\lambda}(X)
	F^a_{\alpha\nu}(X')\rangle
-\langle F^a_{\kappa\nu}(X)F^a_{\alpha\lambda}(X')\rangle
\Big)\partial_{p}^{\kappa}\partial_{p}^{\alpha}f^{\rm s}_{V}(p)\bigg]
+\chi_1+\chi_2
\\\nonumber
&=&-\frac{\bar{C}_2}{2}
\epsilon^{\mu\nu\rho\sigma}\int^{p}_{k,\delta X}
\big(p_{\rho}\langle F^{a\lambda}_{\nu}F^a_{\alpha\lambda}\rangle+p^{\lambda}\langle F^{a}_{\rho\lambda}F^a_{\alpha\nu}\rangle\big)\big(\partial_{\delta X\sigma}\Phi(\delta X )\big)\partial_{p}^{\alpha}f^{\rm s}_{V}(p)
+\chi_1+\chi_2,
\\
\end{eqnarray}
where 
$\int^{p}_{k,X'}=\int^{p}_{k,\delta X}\equiv \int d^4k\int\frac{d^4\delta X}{(2\pi)^4}e^{-ik\cdot\delta X}\big(\pi\delta(p\cdot k)+iPV(1/p\cdot k)\big)$ and we have used $\langle F^a_{[\kappa\nu}F^a_{\alpha]\lambda}\rangle \partial_{p}^{\kappa}\partial_{p}^{\alpha}f^{\rm s}_{V}(p)=0$.
Here
\begin{eqnarray}
\chi_1=-\frac{\mathscr{A}^{\mu}_{0}[f^{\rm s}_{V}(p)]}{p_0}=\frac{\bar{C}_2}{2p_0}
\epsilon^{\mu\nu\rho\sigma}\int^{p}_{k,\delta X}p^{\lambda}p_{\rho}\langle F^a_{0[\lambda}F^a_{\alpha\nu]}\rangle
\big(\partial_{\delta X\sigma}\Phi(\delta X)\big)\partial^{\alpha}_{p}f^{\rm s}_{V}(p)
\end{eqnarray}
and
\begin{eqnarray}
	\chi_2=-\frac{\bar{C}_2}{2p_0}
	\epsilon^{\mu\nu\rho\sigma}\int^{p}_{k,\delta X}\delta X_0p^{\lambda}p_{\rho}\langle F^a_{\kappa[\lambda}F^a_{\alpha\nu]}\rangle
	\big(\partial_{\delta X^{\gamma}_{\perp}}\partial_{\delta X\sigma}\Phi(\delta X)\big)\left[\partial_{p\kappa}\left(\frac{p^{\gamma}_{\perp}}{p_0}\right)\right]\partial^{\alpha}_{p}f^{\rm s}_{V}(p),
\end{eqnarray}
while their contributions will vanish in the following approximations.

We may now introduce the chromo-electric and chromo-magnetic fields explicitly via
\begin{eqnarray}
	F^{a}_{\kappa\lambda}=-\epsilon_{\kappa\lambda\xi\eta}B^{a\xi}\bar{n}^{\eta}+E^a_{[\kappa}\bar{n}_{\lambda]},\quad
	\tilde{F}^{a\mu\kappa}=B^{a[\mu}\bar{n}^{\kappa]}+\epsilon^{\mu\kappa\xi\eta}E^a_{\xi}\bar{n}_{\eta}.
\end{eqnarray} 
When assuming the hierarchy $|\langle B^{a}_{\mu}B^{a}_{\nu}\rangle| \gg |\langle E^{a}_{\mu}B^{a}_{\nu}\rangle| \gg |\langle E^{a}_{\mu}E^{a}_{\nu}\rangle|$, stemming from the screening of chromo-electric field as opposed to the chromo-magnetic field albeit in the static case \cite{Weldon:1982aq} and amplification of the latter from plasma instability in anisotropic QGP \cite{Mrowczynski:1993qm}, the color-singlet SKE is satisfied by $f^{\rm s}_{V}(p)=f_{\rm eq}(p\cdot u)\equiv 1/(e^{\beta (p\cdot u-\mu)}+1)$ in equilibrium with $u^{\mu}$ the fluid four velocity and $\beta=1/T$ the inverse of temperature \footnote{Nonzero $\langle E^{a}_{\mu}B^{a}_{\nu}\rangle$ or $\langle E^{a}_{\mu}E^{a}_{\nu}\rangle$ can further lead to non-equilibrium corrections such that $\delta f^{\rm s}_{V}(p)= f^{\rm s}_{V}(p)-f_{\rm eq}(p\cdot u)\neq 0$, but these corrections will be at $\mathcal{O}(g^2)$ and suppressed at weak coupling.}. For the parity-odd correlation, we further assume the symmetric condition $\langle E^{a}_{\mu}B^{a}_{\nu}\rangle=\langle B^{a}_{\mu}E^{a}_{\nu}\rangle$ for simplification. One could further derive the deviation of $f^{\rm s}_{V}$ near local equilibrium to extract the anomalous transport coefficient from the diffusion term \cite{Asakawa:2006jn}. We may focus on the spin polarization when the charge distribution of quarks reaches local equilibrium with negligible corrections from spacetime gradients of fluid velocity, temperature, and chemical potential. 
Assuming the absence of initial spin polarization in the non-equilibrium phase, the spin diffusion term is then relatively negligible than the source term. The suppression of the diffusion term will be explicitly shown later. Accordingly, Eq.~(\ref{AKE_signlet_simplify}) reduces to
\begin{eqnarray}\label{AKE_singlet_sourceonly}
	0=\delta(p^2-m^2)\Big[p\cdot\partial\tilde{a}^{\rm s\mu}(p,X)
	+\hbar\partial_{p}^{\kappa}\big(\mathscr{A}^{\mu}_{\kappa}[f^{\rm s}_{V}]\big)_{\rm eq}\Big],
\end{eqnarray}
where $\chi_1$ vanishes and we find
\begin{eqnarray}\nonumber
	\partial_{p}^{\kappa}\big(\mathscr{A}^{\mu}_{\kappa}[f^{\rm s}_{V}]\big)_{\rm eq}&\approx &-\frac{\bar{C}_2}{2}
	\int^{p}_{k,\delta X}\big(\partial_{p0}f_{\rm eq}(p_0)\big)\Big[\langle B^{a\mu}E^{a\nu}\rangle p_{\nu}u\cdot \partial_{\delta X}+\langle B^a\cdot E^a\rangle u^{[\mu}p^{\nu]}\partial_{\delta X\nu}
	-\langle B^{a\rho}E^{a\nu}\rangle 
	\\
	&&\times u^{\mu}p_{\nu}\partial_{\delta X\rho}
	+\epsilon^{\mu\nu\rho\sigma}u_{\rho}(\langle E^a_{\nu}E^a_{\lambda}\rangle p^{\lambda}-\langle E^a\cdot E^a\rangle p_{\nu})\partial_{\delta X\sigma}
	\Big]\Phi(\delta X)+\chi_2
\end{eqnarray}
by working in the fluid rest frame (not to be confused with the frame choice for the spin basis) such that $u^{\mu}\approx (1,{\bm u})$ with $|\bm u|\ll 1$.
Here we have taken $u^{\mu}\approx \bar{n}^{\mu}$ 
and introduced the condition $\langle B^{a\mu}(X)E^{a\beta}(X')\rangle=\langle E^{a\beta}(X)B^{a\mu}(X')\rangle$. When further imposing the spatial homogeneity of the field-strength correlators such that $\Phi(\delta X)=\Phi(\delta X_0)$ and accordingly $\chi_2=0$, it is found
\begin{eqnarray}\nonumber
	\partial_{p}^{\kappa}\big(\mathscr{A}^{\mu}_{\kappa}[f^{\rm s}_{V}]\big)_{\rm eq}&=&-\frac{\bar{C}_2}{2}
	\int dk_0\int\frac{d\delta X_0}{2\pi}e^{-ik_0\delta X_0}\Big[\pi\delta(p_0k_0)+i{\rm PV}\Big(\frac{1}{p_0k_0}\Big)\Big]\big(\partial_{p0}f_{\rm eq}(p_0)\big)
	\\\nonumber
	&&\times\big(\langle B^{a\mu}E^{a\nu}\rangle p_{\nu}-\langle B^a\cdot E^a\rangle p^{\mu}_{\perp}\big)\partial_{\delta X_0}\Phi(\delta X_0)
	\\\nonumber
&=&-\frac{\bar{C}_2}{4}
\int d\delta X_0\frac{(1+{\rm sgn}(\delta X_0))}{p_0}\big(\partial_{p0}f_{\rm eq}(p_0)\big)\big(\langle B^{a\mu}E^{a\nu}\rangle p_{\nu}-\langle B^a\cdot E^a\rangle p^{\mu}_{\perp}\big) 
\\
&&\times \partial_{\delta X_0}\Phi(\delta X_0).
\end{eqnarray}
Notably, when $\Phi(x)|_{x\rightarrow \pm\infty}=0$, one can carry out the integration by parts for the equation above and derive 
\begin{eqnarray}
\partial_{p}^{\kappa}\big(\mathscr{A}^{\mu}_{\kappa}[f^{\rm s}_{V}]\big)_{\rm eq}=\frac{\bar{C}_2\Phi(0)}{2p_0}
\big(\partial_{p0}f_{\rm eq}(p_0)\big)\big(\langle B^{a\mu}E^{a\nu}\rangle p_{\nu}-\langle B^a\cdot E^a\rangle p^{\mu}_{\perp}\big).
\end{eqnarray}

Setting 
\begin{eqnarray}
\Phi(\delta X_0)=e^{-\delta X_0^2/\tau^2_{c}}
\end{eqnarray}
as the Gaussian form with $\tau_c$ the correlation time, we further obtain 
\begin{eqnarray}\label{sorce_simpl}
	\partial_{p}^{\kappa}\big(\mathscr{A}^{\mu}_{\kappa}[f^{\rm s}_{V}]\big)_{\rm eq}
	=\frac{\bar{C}_2}{2p_0}
	\big(\partial_{p0}f_{\rm eq}(p_0)\big)\big(\langle B^{a\mu}E^{a\nu}\rangle p_{\nu}-\langle B^a\cdot E^a\rangle p^{\mu}_{\perp}\big).
\end{eqnarray}
Since now $\partial_{p}^{\kappa}\big(\mathscr{A}^{\mu}_{\kappa}[f^{\rm s}_{V}]\big)_{\rm eq}$ is independent of $X$, $\tilde{a}^{\rm s\mu}$ will be a function linear to time. More precisely, by solving Eq.~(\ref{AKE_singlet_sourceonly}) with Eq.~(\ref{sorce_simpl}), we obtain \footnote{The $\langle B^{a\mu}E^{a\nu}\rangle p_{\nu}$ term below will be in fact $\langle E^{a\mu}B^{a\nu}\rangle p_{\nu}$ without using the symmetric correlation $\langle B^{a\mu}E^{a\nu}\rangle=\langle B^{a\nu}E^{a\mu}\rangle$.}
\begin{eqnarray}\label{amu_from_source}\nonumber
\tilde{a}^{\rm s\mu}(t,p)&=&-\frac{\hbar\bar{C}_2(t-t_0)}{2p_0^2}\Phi(0)\big(\partial_{p0}f_{\rm eq}(p_0)\big)\big(\langle B^{a\mu}E^{a\nu}\rangle p_{\nu}-\langle B^a\cdot E^a\rangle p^{\mu}_{\perp}\big)
\\
&=&-\frac{\hbar\bar{C}_2(t-t_0)}{2p_0^2}
\big(\partial_{p0}f_{\rm eq}(p_0)\big)\big(\langle B^{a\mu}E^{a\nu}\rangle p_{\nu}-\langle B^a\cdot E^a\rangle p^{\mu}_{\perp}\big),
\end{eqnarray}
for $\tilde{a}^{\mu}(t_0,p)=0$. This secular solution stems from only the $\delta X$ dependence of color-field correlators. Notwithstanding the ostensible $\tau_{c}$ independence of Eq.~(\ref{sorce_simpl}), the source term should vanish when color-field correlators become constant. That is, $\partial_{p}^{\kappa}\big(\mathscr{A}^{\mu}_{\kappa}[f^{\rm s}_{V}]\big)_{\rm eq}\rightarrow 0$ when $\tau_c\rightarrow \infty$. In a finite system, there exists an upper bound for $|\delta X|$ determined by the size of the system. When $\tau_c$ is greater than the system size, $\Phi(\delta X_0)$ is peaked at $\delta X_0=0$ and the source term in the AKE is suppressed. In heavy ion collisions, one may instead assume an approximately infinite system. 

In principle, at a sufficiently long time, $\tilde{a}^{s\mu}$ is no longer small and the diffusion term should start to play a role. According to the proposed hierarchy, the diffusion term in the color-singlet AKE is given by  
\begin{eqnarray}\nonumber
\partial_{p}^{\kappa}\mathscr{D}_{\kappa}[\tilde{a}^{s\mu}]&\approx&\bar{C}_2\epsilon_{\alpha\beta\gamma\kappa}\epsilon_{\nu\rho\lambda\sigma}u^{\alpha}u^{\nu}\partial_{p}^{\beta}\int^p_{k,X'}\langle B^{a\kappa}(X)B^{a\sigma}(X')\rangle p^{\gamma}p^{\lambda}\partial_{p}^{\rho}\tilde{a}^{s\mu}
\\
&=&\frac{\bar{C}_2\sqrt{\pi}\tau_m}{2p_0}\epsilon_{\alpha\beta\gamma\kappa}\epsilon_{\nu\rho\lambda\sigma}u^{\alpha}u^{\nu}\langle B^{a\kappa}B^{a\sigma}\rangle\partial_{p}^{\beta}\big( p^{\gamma}p^{\lambda}\partial_{p}^{\rho}\tilde{a}^{s\mu}\big),
\end{eqnarray}
where we derive the second equality by assuming $\langle B^{a\kappa}(X)B^{a\sigma}(X')\rangle=\langle B^{a\kappa}B^{a\sigma}\rangle e^{-\delta X_0^2/\tau_{m}^2}$. Conducting a simple power counting, $\partial_{p}^{\kappa}\mathscr{D}_{\kappa}[\tilde{a}^{s\mu}]$ becomes comparable to $p\cdot \partial \tilde{a}^{s\mu}$ when
\begin{eqnarray}
(t-t_0)\tau_m|\langle B^{a\kappa}B^{a\sigma}\rangle|\sim p_0^2.
\end{eqnarray} 
Based on the kinetic region we considered for the $\hbar$ expansion, $p_0^2\gg |B^{a\mu}|$, we may in general neglect the diffusion term given no other sources for spin polarization. Nevertheless, the suppression of the diffusion term is also valid because of the properties of postulated color fields, which may not be always the case in a more general condition.

Moreover, nonzero spin polarization could be engendered by the explicit $\hbar$ term in Eq.~(\ref{eq:Asmu_massive}), which could be regarded as a non-dynamical source term for spin polarization, even when the dynamical source term in the AKE is suppressed. We will analyze such a contribution from Eq.~(\ref{eq:Asmu_massive}) in the following paragraphs. When $f^{\rm s}_{V}(p)$ is in thermal equilibrium, one obtains
\begin{eqnarray}\label{AQ1_massive}
\mathcal{A}_{Q1}^{\mu}	= -\frac{1}{2}\bigg[\int^{p}_{k,X'}\langle \tilde{F}^{a\mu\kappa}(X)E^a_{\beta}(X')\rangle \big(\delta^{\beta}_{\kappa}+u_{\kappa}(p^{\beta}\partial_{0}-\hat{p}^{\beta})\big)\partial_{p0}f_{\rm eq}(p_0)\bigg]_{p_0=\epsilon_{\bm p}}
+\xi_1
\end{eqnarray} 
and
\begin{eqnarray}\nonumber
\mathcal{A}^{\mu}_{Q2}&=&-\frac{1}{2\epsilon_{\bm p}^2}(p_{\perp\kappa}-\epsilon_{\bm p}^2\partial_{p_{\perp}\kappa})\bigg[\int^{p}_{k,X'}p^{\beta}\langle \tilde{F}^{a\mu\kappa}(X)E^a_{\beta}(X')\rangle \partial_{p0}f_{\rm eq}(p_0)\bigg]_{p_0=\epsilon_{\bm p}}+\xi_2
\\		
&=&-\frac{1}{2}\bigg[\int^{p}_{k,X'}\langle \tilde{F}^{a\mu\kappa}(X)E^a_{\beta}(X')\rangle \big(\hat{p}_{\perp\kappa}\hat{p}^{\beta}(2-p_0\partial_{p0})-\delta^{\beta}_{\kappa}\big)\partial_{p0}f_{\rm eq}(p_0)\bigg]_{p_0=\epsilon_{\bm p}}+\xi_2
\end{eqnarray}
Here 
\begin{eqnarray}
	\xi_1= -\frac{1}{2}\bigg[\int^{p}_{k,X'}\delta X_0\partial_{\delta X_{\perp}\rho}\langle \tilde{F}^{a\mu\kappa}(X)E^a_{\beta}(X')\rangle p^{\beta}\big(\partial_{p0}f_{\rm eq}(p_0)\big)\partial_{p\kappa}\left(\frac{p^{\rho}_{\perp}}{p_0}\right)\bigg]_{p_0=\epsilon_{\bm p}}
\end{eqnarray}
and
\begin{eqnarray}
\xi_2=\frac{1}{2}\bigg[\int^{p}_{k,X'}\delta X_0\partial_{\delta X_{\perp}\rho}\langle \tilde{F}^{a\mu\kappa}(X)E^a_{\beta}(X')\rangle p^{\beta}\partial_{p0}f_{\rm eq}(p_0)\bigg]_{p_0=\epsilon_{\bm p}}\partial_{p_{\perp}\kappa}\left(\frac{p^{\rho}_{\perp}}{\epsilon_{\bm p}}\right).
\end{eqnarray}

Further taking $\langle E^{a\mu}(X) B^{a\beta}(X')\rangle=\langle E^{a\mu}(X') B^{a\beta}(X)\rangle$, $\langle E^{a\mu}(X) E^{a\beta}(X')\rangle=\langle E^{a\mu}(X') E^{a\beta}(X)\rangle$, and $\Phi(\delta X)=\Phi(\delta X_0)$ such that $\xi_1=\xi_2=0$
and using the relation,
\begin{eqnarray}\label{EB_rel}\nonumber
	\tilde{F}^{a\mu\beta}(X)F^a_{\alpha\beta}(X')&=&\delta^{\mu}_{\alpha}E^a(X)\cdot B^a(X')+\bar{n}^{\mu}\bar{n}_{\alpha}\big(B^a(X)\cdot E^a(X')-E^a(X)\cdot B^a(X')\big)
	\\	&&+B^{a\mu}(X)E^a_{\alpha}(X')-E^a_{\alpha}(X)B^{a\mu}(X'),
\end{eqnarray}  
with the symmetric correlation $\langle E^{a\mu}(X) B^{a\beta}(X')=\langle E^{a\beta}(X) B^{a\mu}(X')\rangle$,
it is found
\begin{eqnarray}\nonumber
\mathcal{A}_{Q1}^{\mu}(\bm p,X)&\approx&\bigg[\int d\delta X_0\frac{(1+{\rm sgn}(\delta X_0))}{4p_0}\Phi(\delta X_0)\big(\langle E^a\cdot B^a\rangle u^{\mu}-\langle B^{a\mu}E^{a\beta}\rangle (p_{\beta}\partial_{p_0}-\hat{p}_{\beta})\big)
\\
&&\times\partial_{p_0}f_{\rm eq}(p_0)\bigg]_{p_0=\epsilon_{\bm p}},
\end{eqnarray}
and
\begin{eqnarray}\nonumber
	\mathcal{A}_{Q2}^{\mu}(\bm p,X)&\approx&
	\bigg[\int d\delta X_0\frac{(1+{\rm sgn}(\delta X_0))}{4p_0}\Phi(\delta X_0)
	\Big(\big(\hat{p}^{\alpha}\hat{p}^{\beta}\langle E^a_{\alpha}B^a_{\beta}\rangle u^{\mu}+\epsilon^{\nu\mu\rho\alpha}u_{\nu}\hat{p}_{\rho}\hat{p}^{\beta}\langle E^a_{\alpha}E^a_{\beta}\rangle\big)
	\\
&&\times (2-p_0\partial_{p0})-\langle E^a\cdot B^a\rangle u^{\mu} \Big)\partial_{p_0}f_{\rm eq}(p_0)\bigg]_{p_0=\epsilon_{\bm p}},
\end{eqnarray}
which yield 
\begin{eqnarray}\nonumber\label{AQ_massive}
	\mathcal{A}_Q^{\mu}(\bm p,X)&=&\frac{\pi^{1/2}\tau_c}{4\epsilon_{\bm p}^3}\Big[\big(p^{\alpha}p^{\beta}\langle E^a_{\alpha}B^a_{\beta}\rangle u^{\mu}+\epsilon^{\mu\alpha\rho\nu}u_{\nu}p_{\rho}\langle E^a_{\alpha}E^a_{\beta}\rangle\big)(2p^{\beta}-\epsilon_{\bm p}^2\partial^{\beta}_{p})
	\\
	&&-\langle B^{a\mu}E^{a\beta}\rangle \epsilon_{\bm p}(\epsilon_{\bm p}^2\partial_{p\beta}-p_{\beta})\Big]\partial_{\epsilon_{\bm p}}f_{\rm eq}(\epsilon_{\bm p})
\end{eqnarray}
for $\Phi(\delta X_0)=e^{-\delta X_0^2/\tau^2_{c}}$, where we have used $\epsilon_{\bm p}\partial^{\beta}_{p_{\perp}}G(\epsilon_{\bm p})= p^{\beta}_{\perp}\partial_{\epsilon_{\bm p}}G(\epsilon_{\bm p})$ for an arbitrary function $G(\epsilon_{\bm p})$. 
In $\mathcal{A}_{Q2}^{\mu}$, we further have nonzero contribution from $\langle E^{a}_{\nu}E^{a\beta}\rangle$ as a parity-even correlator, while it is attached to $\epsilon^{\mu\alpha\rho\nu}u_{\nu}\hat{p}_{\rho}$ and hence the overall parity-odd contribution is induced. Such a term could be understood as the combination of the Lorentz force and spin Hall effect, whereas it is subdominant according to our assumption that the average magnitude of color magnetic fields is larger than of color electric fields.  This term also vanishes when assuming $\langle E^a_{\alpha}E^a_{\beta}\rangle \sim \delta_{\alpha\beta}$ for the correlation of only parallel color fields. When combining the contribution from $\tilde{a}^{\rm s\mu}$ in Eq.~(\ref{amu_from_source}), one finds 
\begin{eqnarray}\nonumber\label{Amu_equil}
	\hat{\mathcal{A}}^{\rm s\mu}(\bm p,X)
	&=&\frac{\hbar\bar{C}_2}{4\epsilon_{\bm p}^3}\Big\{\sqrt{\pi}\tau_c\Big[(p^{\alpha}p^{\beta}\langle E^a_{\alpha}B^a_{\beta}\rangle u^{\mu}+\epsilon^{\mu\alpha\rho\nu}u_{\nu}p_{\rho}\langle E^a_{\alpha}E^a_{\beta}\rangle)(2p^{\beta}-\epsilon_{\bm p}^2\partial^{\beta}_{p})
	-\langle B^{a\mu}E^{a\nu}\rangle
	\\
	&&\times \epsilon_{\bm p}(\epsilon_{\bm p}^2\partial_{p\nu}-p_{\nu})\Big]
	+2(t-t_0)\epsilon_{\bm p}\big(\langle E^a\cdot B^a\rangle p^{\mu}_{\perp}-\langle B^{a\mu}E^{a\nu}\rangle p_{\nu}\big)
	\Big\}\partial_{\epsilon_{\bm p}}f_{\rm eq}(\epsilon_{\bm p}).
\end{eqnarray}
When considering the non-relativistic limit such that $|p^{\mu}_{\perp}|\ll \epsilon_{\bm p}$, we further obtain 
\begin{eqnarray}
\hat{\mathcal{A}}^{\rm s\mu}(\bm p,X)\approx -\frac{\hbar\bar{C}_2}{4\epsilon_{\bm p}}\sqrt{\pi}\tau_c\langle B^{a\mu}E^{a\nu}\rangle p_{\nu}\partial_{\epsilon_{\bm p}}f_{\rm eq}(\epsilon_{\bm p}),
\end{eqnarray}
where one component of the non-dynamical source term dominates. Such a term stems from the combination of spin polarization induced by a color magnetic field and the Lorentz force driven by a color electric field. We may further evaluate the axial charge current via Eq.~(\ref{def_currents}), where only $\mathcal{A}^{\mu}_{Q1}$ contributes. One hence obtains 
\begin{eqnarray}\label{J5_massive}
	J^{\mu}_5
	=2\hbar \int\frac{d^4p}{(2\pi)^3}\frac{\delta(p^2-m^2)}{4p\cdot u}\sqrt{\pi}\tau_c
	\langle E^a\cdot B^a\rangle u^{\mu}\partial_{p\cdot u}f_{\rm eq}(p\cdot u)
	.
\end{eqnarray}
Here the contribution from $\tilde{a}^{\rm s\mu}$ in Eq.~(\ref{amu_from_source}) does not affect $J^{\mu}_5$. 

\subsection{Axial Ward identity}
\begin{figure}
	\begin{center}
		\includegraphics[width=0.6\hsize]{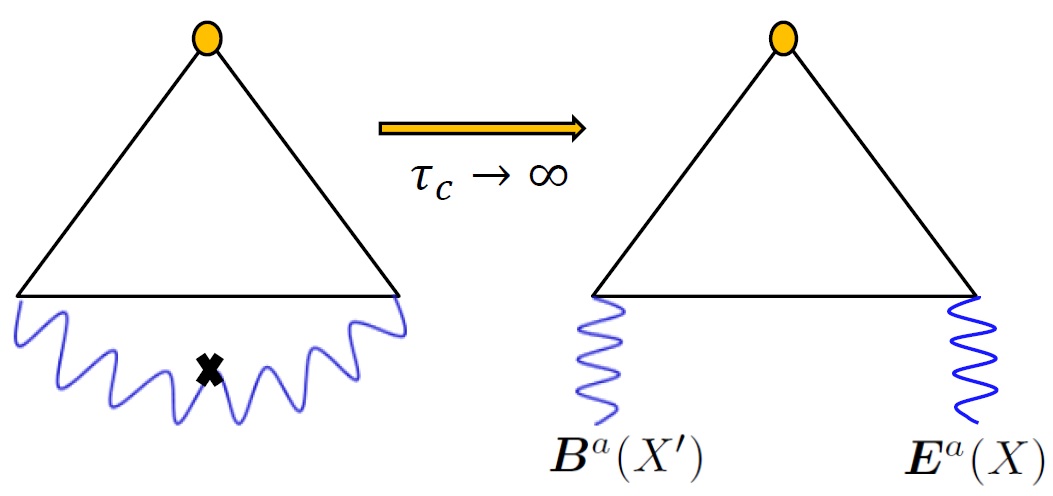}
	\end{center}
	\caption{The triangle diagram with finite $\tau_c$ and with $\tau_c\rightarrow \infty$ in the constant-field limit. The blobs denote the axial vertices and the wiggly line with a cross at the center represents the correlator of color fields originating from the medium.
	}
	\label{fig:triangle diagrams}
\end{figure}

It is obvious that $J^{\mu}_5$ in Eq.~(\ref{J5_massive}) is independent of $X$ and thus $\partial_{\mu}J^{\mu}_5=0$.
It is however useful to investigate the vanishing axial Ward identity from Eq.~(\ref{AQ1_massive}).
When $\langle B^{a}(X)\cdot E^a(X')\rangle$ only depends on $\delta X$, it is found
\begin{eqnarray}\nonumber\label{eq:partialx_EB}
	\partial_{X\mu}\int^{p}_{k,X'}
	\langle B^{a}(X)\cdot E^a(X')\rangle&=&
	\int^{p}_{k,\delta X}
	(-ik_{\mu}+\partial_{\delta X\mu})\langle B^{a}(X)\cdot E^a(X')\rangle
	\\
	&=&i\int d^4k\frac{d^4\delta X}{(2\pi)^4}e^{-ik\cdot\delta X}\frac{k_{\mu}+i\partial_{\delta X\mu}}{p\cdot k+i\epsilon}\langle B^{a}(X)\cdot E^a(X')\rangle.
\end{eqnarray}
From the integration by part, $i\partial_{\delta X\mu}$ in the integrand above could be replaced by $-k_{\mu}$ and thus we find $\partial_{\mu}J^{\mu}_5(X)=0$ for finite $\tau_c$. 
Nevertheless, when $\tau_c\rightarrow \infty$ such that $\langle B^{a}(X)\cdot E^a(X')\rangle\rightarrow \langle B^{a}\cdot E^a\rangle$ as the constant-field limit, $i\partial_{\delta X\mu}$ in the integrand of Eq.~(\ref{eq:partialx_EB}) no longer contributes and we find
\begin{eqnarray}
	\partial_{\mu} J_5^{\mu}(X)=2\hbar \int d^4k\frac{d^4\delta X}{(2\pi)^4}e^{-ik\cdot\delta X}\langle B^{a}\cdot E^a\rangle\int\frac{d^4p}{(2\pi)^3}{{\rm sgn}(p_0)}\frac{\delta(p^2-m^2)}{2(p\cdot k+i\epsilon)}
 k_0\partial_{p_0}f_{\rm eq}(p_0),
\end{eqnarray}
where we further insert ${{\rm sgn}(p_0)}$ as a sign function for energy to include the contribution from anti-quarks. In fact, the ultraviolet contribution from anti-fermions is essential to reproduce the term associated with the chiral anomaly \cite{Hidaka:2017auj}. Now, the $\delta X$ integral can be evaluated independently,
\begin{eqnarray}\label{eq:delta_k}
	\int \frac{d^4\delta X}{(2\pi)^4}e^{ik\cdot\delta X}=\delta^4(k).
\end{eqnarray}
For the remaining $p$ and $k$ integrals, one may make the decomposition,
\begin{eqnarray}
	\int d^4k\delta^4(k)\int\frac{d^4p}{(2\pi)^3}{{\rm sgn}(p_0)}\frac{\delta(p^2-m^2)}{2(p\cdot k+i\epsilon)}k_0\partial_{p_0}f_{\rm eq}(p_0)
	=I_1+I_2,
\end{eqnarray}
where the real part of $(p\cdot k+i\epsilon)^{-1}$ contributes to \footnote{Here we first take $\lim_{\bm k\rightarrow 0}$ and then $\lim_{k_0\rightarrow 0}$. In fact, for constant color fields, we may reduce $\int^{p}_{k,X'}$ to just the one-dimensional integral as $\int^{p}_{k,X'}\equiv \int dk_0\int\frac{dX'_0}{(2\pi)}e^{ik_0(X'_0-X_0)}i(p_0k_0+i\epsilon)^{-1}$ and hence the $\bm k$ and $\bm X'$ integrals are redundant since it is natural to assume $f^a_{V}$ and $\tilde{a}^{\mu}_a$ only depend on time.}
\begin{eqnarray}
	I_1=\int\frac{d^4p}{(2\pi)^3}{{\rm sgn}(p_0)}\frac{\delta(p^2-m^2)}{2p\cdot k}k_0\partial_{p_0}f_{\rm eq}(p_0)\Big|_{k_{\mu}\rightarrow 0}=\int\frac{d^3\bm p}{(2\pi)^3 4\epsilon_{\bm p}^2}\frac{d}{d\epsilon_{\bm p}}\big[f_{\rm eq}(\epsilon_{\bm p})-f_{\rm eq}(-\epsilon_{\bm p})\big]
\end{eqnarray}
with $\epsilon_{\bm p}=\sqrt{|\bm p|^2+m^2}$, and the imaginary part of $(p\cdot k+i\epsilon)^{-1}$ yields
\begin{eqnarray}
	I_2&=&-i\int\frac{d^4p}{(2\pi)^3}{{\rm sgn}(p_0)}\frac{\delta(p^2-m^2)}{2}k_0\pi\delta(p\cdot k)\partial_{p_0}f_{\rm eq}(p_0)\Big|_{k_{\mu}\rightarrow 0}=0.
\end{eqnarray}
In the end, we obtain an axial Ward identity as 
\begin{eqnarray}\label{divJ5_nonzero}
	\partial_{\mu} J_5^{\mu}(X)=\hbar \langle B^{a}\cdot E^a\rangle
	\int^{\infty}_0\frac{d|\bm p||\bm p|^2}{4\pi^2\epsilon_{\bm p}^2}\frac{d}{d\epsilon_{\bm p}}\big[f_{\rm eq}(\epsilon_{\bm p})-f_{\rm eq}(-\epsilon_{\bm p})\big].
\end{eqnarray} 
As briefly discussed in Ref.~\cite{Muller:2021hpe}, the vanishing axial Ward identity at finite $\tau_c$ corresponds to a triangle diagram with two gluon legs connected and zero momentum flow from the axial vertex. However, when $\tau_c\rightarrow \infty$, two legs break apart and allow the momentum exchanges. See Fig.~\ref{fig:triangle diagrams} for a schematic description. 

It is found $\partial_{\mu} J^{\mu}_5(X)\neq 0$ as opposed to the conservation of a vector charge current, $\partial_{\mu} J^{\mu}_V(X)=0$, implied by Eq.~(\ref{SKE_signlet_simplify}) since the diffusion term can be recast into a total momentum derivative including the attached $\delta(p^2-m^2)$ due to $p^{\kappa}\mathscr{D}_{\kappa}[O]=0$. Now, Eq.~(\ref{divJ5_nonzero}) seems to be in connection to the chiral anomaly but it also receives the finite-mass correction. Recall that 
\begin{eqnarray}
	\partial_{\mu} J^{\mu}_5(X)=-\frac{E\cdot B}{2\pi^2}+2m\langle \bar{\psi}i\gamma_5\psi\rangle
\end{eqnarray} 
for massive fermions, where we consider the Abelian gauge fields for simplicity yet the similar feature holds for non-Abeleian cases. Here the pseudo-scalar condensate results in further modifications when $m\neq 0$. However, it is anticipated that $\partial_{\mu} J^{\mu}_5(X)=0$ when $m\rightarrow \infty$. For example, in the vacuum with constant electromagnetic fields, it is found \cite{Copinger:2018ftr},
\begin{eqnarray}
	\langle \bar{\psi}i\gamma_5\psi\rangle=\frac{E\cdot B}{4\pi^2m}\Big(1-e^{-\pi m^2/|E|}\Big)
\end{eqnarray} 
and hence
\begin{eqnarray}
	\partial_{\mu} J^{\mu}_5(X)=-\frac{E\cdot B}{2\pi^2}e^{-\pi m^2/|E|}.
\end{eqnarray}
As shown in Eq.~(\ref{divJ5_nonzero}), $\partial_{\mu} J^{\mu}_5(X)=0$ when $m\rightarrow \infty$. In fact, from Eq.~(\ref{divJ5_nonzero}), we can extract the pseudo-scalar condensate,
\begin{eqnarray}
\langle \bar{\psi}i\gamma_5\psi\rangle=-\frac{\hbar \langle B^{a}\cdot E^a\rangle}{8m\pi^2}
\int^{\infty}_0d|\bm p|\left(1-\frac{|\bm p|}{\epsilon_{\bm p}}\right)\frac{d}{d|\bm p|}\big[f_{\rm eq}(\epsilon_{\bm p})-f_{\rm eq}(-\epsilon_{\bm p})\big],
\end{eqnarray}
such that Eq.~(\ref{divJ5_nonzero}) is decomposed as
\begin{eqnarray}\label{divJ5_condensate}
\partial_{\mu} J_5^{\mu}(X)=-\hbar \frac{\langle B^{a}\cdot E^a\rangle}{4\pi^2}+2m\langle \bar{\psi}i\gamma_5\psi\rangle,
\end{eqnarray}
where the first term on the right-hand side above is given by
\begin{eqnarray}\nonumber
&&\frac{\hbar}{2}\langle B^{a}\cdot E^a\rangle
\int^{\infty}_0\frac{d|\bm p|}{4\pi^2}\frac{d}{d|\bm p|}\big[f_{\rm eq}(|\bm p|)-f_{\rm eq}(-|\bm p|)\big]
\\\nonumber
&&=\frac{\hbar}{4\pi^2}\langle B^{a}\cdot E^a\rangle
\big[f_{\rm eq}(|\bm p|)-f_{\rm eq}(-|\bm p|)\big]\Big|^{|\bm p|=\infty}_{|\bm p|=0}
\\
&&=-\frac{\hbar}{4\pi^2}\langle B^{a}\cdot E^a\rangle.
\end{eqnarray}
Naively, when taking $m\rightarrow 0$, it seems $m\langle \bar{\psi}i\gamma_5\psi\rangle\rightarrow 0$ and the non-Abelian axial anomaly is reproduced in the massless limit. 
Nevertheless, since we work in the rest frame $n^{\mu}=n^{\mu}_{r}(p)$, the result is in principle subject to a finite-mass regime $m^2\gg |E^a|,\,|B^a|$. We may have to examine the massless limit more rigorously as will be demonstrated in the follow-up section. 

\section{Massless fermions}\label{sec:massless_fermions}
In order to investigate the influence of chromo-electromagnetic fields upon light quarks in QGP. We further analyze the similar effects on spin polarization of massless quarks via the CKT with non-Abelian gauge fields to complement the previous study for massive quarks.   

\subsection{Derivation of the effective AKE}
For massless fermions, it is more convenient to work in the chiral bases and the master equations reduce to 
\begin{eqnarray}
	&&\hat{\nabla}\cdot \mathcal{W}_{s}=0,
	\\
	&&\hat{\Pi}\cdot\mathcal{W}_{s}=0,
	\\
	&&\hat{\Pi}_{\nu}\mathcal{W}_{s\mu}-\hat{\Pi}_{\mu}\mathcal{W}_{s\nu}
	=\frac{s\hbar}{2}\epsilon_{\mu\nu\rho\sigma}\hat{\nabla}^{\rho}\mathcal{W}_{s}^{\sigma},
\end{eqnarray}
where $\mathcal{W}_{s\mu}(p,X)$ denote the chiral components of Wigner operators (lesser propagators) for massless fermions with $s=1$ and $s=-1$ for right and left handed fermions (not to confuse with the superscript $\rm s$ for color singlet). Unlike the case for Dirac fermions, $\mathcal{W}_{s}$ are $2\times 2$ matrices in spinor space, which can be decomposed as $\mathcal{W}_{+}=\bar{\sigma}^{\mu}\mathcal{W}_{+\mu}$ and $\mathcal{W}_{-}=\sigma^{\mu}\mathcal{W}_{-\mu}$, where $\sigma^{\mu}=(1,{\bm \sigma})$ and $\bar{\sigma}^{\mu}=(1,-{\bm \sigma})$ with $\sigma^i$ being Pauli matrices. Based on the $\hbar$ expansion, the master equations explicit read
\begin{eqnarray}\label{Weyl_eq_1}
	&&D\cdot \mathcal{W}_{s}+\frac{1}{2}\{F_{\nu\mu},\partial_{p}^{\nu}\mathcal{W}_{s}^{\mu}\}_{\rm c}-\frac{i\hbar}{24}[(\partial_{p}\cdot DF_{\nu\mu}),\partial_p^{\nu}\mathcal{W}_{s}^{\mu}]_{\rm c}=0,
	\\\label{Weyl_eq_2}
	&&p_{\mu}\mathcal{W}_{s}^{\mu}+\frac{i\hbar}{8}[F_{\nu\mu},\partial_{p}^{\nu}\mathcal{W}_{s}^{\mu}]_{\rm c}=0,
	\\\label{Weyl_eq_3}
	&&p_{[\nu}\mathcal{W}_{s\mu]}+\frac{i\hbar}{8}[F_{\rho[\nu},\partial_{p}^{\rho}\mathcal{W}_{s\mu]}]_{\rm c}
	=\frac{s\hbar}{2}\epsilon_{\mu\nu\rho\sigma}\Big(D^{\rho}\mathcal{W}_{s}^{\sigma}+\frac{1}{2}\{F_{\nu\mu},\partial_{p}^{\nu}\mathcal{W}_{s}^{\sigma}\}_{\rm c}\Big),
\end{eqnarray}
up to $\mathcal{O}(\hbar)$. By contracting Eq.~(\ref{Weyl_eq_3}) with $n^{\mu}$, one can derive the Wigner operator satisfying the constrain in Eq.~(\ref{Weyl_eq_2}). It is found
\begin{eqnarray}\nonumber
	\mathcal{W}_{s}^{\mu}&=&2\pi\bigg[\delta(p^2)\Big(p^{\mu}\hat{f}_s+s\hbar S^{\mu\nu}_{(n)}\tilde{\Delta}_{\nu}\hat{f}_s\Big)
	+\frac{s\hbar}{2} p_{\nu}\delta'(p^2)\{\tilde{F}^{\mu\nu},\hat{f}_s\}_{\rm c}
	+\frac{i\hbar}{8}\bigg(p^{\mu}\delta'(p^2)p^{\rho}[F_{\nu\rho},\partial_{p}^{\nu}\hat{f}_s]_{\rm c}
	\\
	&&+[F^{\rho\mu},\partial_{p\rho}\delta(p^2)\hat{f}_s]_{\rm c}-\frac{2\delta(p^2)}{p\cdot n}[F^{\mu\nu}n_{\nu},\hat{f}_s]_{\rm c}\bigg)\bigg]
\end{eqnarray} 
up to $\mathcal{O}(\hbar)$, where $S^{\mu\nu}_{(n)}=S^{\mu\nu}_{m(n)}|_{m=0}$ represents the spin tensor for massless fermions. When making a conversion to the vector and axial-vector basis by $(\mathcal{V}/\mathcal{A})^{\mu}=(\mathcal{W}^{\mu}_{+}\pm \mathcal{W}^{\mu}_{-})/2$, one finds
\begin{eqnarray}\nonumber
	\mathcal{A}^{\mu}&=&2\pi\bigg[\delta(p^2)\Big(p^{\mu}\hat{f}_{A}+\hbar S^{\mu\nu}_{(n)}\tilde{\Delta}_{\nu}\hat{f}_{V}\Big)
	+\frac{\hbar}{2} p_{\nu}\delta'(p^2)\{\tilde{F}^{\mu\nu},\hat{f}_{V}\}_{\rm c}
	+\frac{i\hbar}{8}\bigg(p^{\mu}\delta'(p^2)p^{\rho}[F_{\nu\rho},\partial_{p}^{\nu}\hat{f}_{A}]_{\rm c}
	\\
	&&+[F^{\rho\mu},\partial_{p\rho}\delta(p^2)\hat{f}_{A}]_{\rm c}-\frac{2\delta(p^2)}{p\cdot n}[F^{\mu\nu}n_{\nu},\hat{f}_{A}]_{\rm c}\bigg)\bigg],
\end{eqnarray}  
where $\hat{f}_{V/A}=(\hat{f}_{+}\pm \hat{f}_{-})/2$. Notably, the $\hbar$ terms related to commutators in color space are all linear to $\hat{f}_A$. As a result, implementing the power counting $\hat{f}_{V}\sim \mathcal{O}(\hbar^0)$ and $\hat{f}_{A}\sim \mathcal{O}(\hbar)$, such terms involving commutators are suppressed. Similarly, the contribution from the last term on the left hand side of Eq.~(\ref{Weyl_eq_1}) in AKE is now of $\mathcal{O}(\hbar^2)$ and hence omitted. That is, the effective AKE is given by 
\begin{eqnarray}\label{AKE_massless}
	\tilde{\Delta}\cdot\mathcal{A}=D\cdot \mathcal{A}+\frac{1}{2}\{F_{\nu\mu},\partial_{p}^{\nu}\mathcal{A}^{\mu}\}_{\rm c}=0,
\end{eqnarray}    
with 
\begin{eqnarray}\label{Amu_massless}
	\mathcal{A}^{\mu}=2\pi\bigg[\delta(p^2)\Big(p^{\mu}\hat{f}_{A}+\hbar S^{\mu\nu}_{(n)}\tilde{\Delta}_{\nu}\hat{f}_{V}\Big)
	+\frac{\hbar}{2} p_{\nu}\delta'(p^2)\{\tilde{F}^{\mu\nu},\hat{f}_{V}\}_{\rm c}
	\bigg],
\end{eqnarray}
which corresponds to the massless limit of Eq.~(\ref{axial_sol}) with $\hat{a}^{\mu}=p^{\mu}\hat{f}_{A}$ due to the spin locking by helicity.  
Note that the $\mathcal{A}^{\mu}$ in Eq.~(\ref{Amu_massless}) is consistent with the result in Ref.~\cite{Luo:2021uog} \footnote{There are extra $\hbar$ terms proportional to $f_A$ in Ref.~\cite{Luo:2021uog}, while these terms are at $\mathcal{O}(\hbar^2)$ with our power counting and accordingly omitted.}. As opposed to the massive case, one is now unable to choose the particle rest frame for $n^{\mu}$. Consequently, the magnetization-current term pertinent to $S^{\mu\nu}_{(n)}$ or the so-called side-jump term \cite{Chen:2014cla,Hidaka:2016yjf} in the massless limit will be always involved. For $\mathcal{V}^{\mu}$, we simply retain the leading-order term of $\mathcal{O}(\hbar^0)$ and thus have $\mathcal{V}^{\mu}=2\pi\delta(p^2)p^{\mu}\hat{f}_V$. Also, the SKE is same as Eq.~(\ref{SKE}) with the massless on-shell condition.

Inserting Eq.~(\ref{Amu_massless}) into Eq.~(\ref{AKE_massless}), the AKE turns out to be
\begin{eqnarray}\nonumber
	0&=&\delta(p^2)\Big(p\cdot\tilde{\Delta}\hat{f}_A+\hbar\big(\tilde{\Delta}_{\mu}S^{\mu\nu}_{(n)}\big)\tilde{\Delta}_{\nu}\hat{f}_V
	+\hbar S^{\mu\nu}_{(n)}\tilde{\Delta}_{\mu}\tilde{\Delta}_{\nu}\hat{f}_V\Big)
	+\hbar \delta'(p^2)S^{\mu\nu}_{(n)}p^{\rho}\{F_{\rho\mu},\tilde{\Delta}_{\nu}\hat{f}_V\}
	\\
	&&+\frac{\hbar}{2}\delta'(p^2) 
	p_{\nu}\tilde{\Delta}_{\mu}\{\tilde{F}^{\mu\nu},\hat{f}_{V}\}_{\rm c}+\frac{\hbar}{4}\big(\partial_{p}^{\rho}p_{\nu}\delta'(p^2)\big)\{F_{\rho\mu},\{\tilde{F}^{\mu\nu},\hat{f}_{V}\}\}_{\rm c}\Big).
\end{eqnarray}
We may now compute each term above. For the $\hbar$ corrections pertinent to $\delta(p^2)$, it is found
\begin{eqnarray}
	\big(\tilde{\Delta}_{\mu}S^{\mu\nu}_{(n)}\big)\tilde{\Delta}_{\nu}\hat{f}_V=(\partial_{\mu}S^{\mu\nu}_{(n)})\tilde{\Delta}_{\nu}\hat{f}_V+\frac{1}{2p\cdot n}\Big(\{\tilde{F}^{\nu\beta}n_{\beta},\tilde{\Delta}_{\nu}\hat{f}_V\}_{\rm c}-S^{\mu\nu}_{(n)}\{n^{\rho}F_{\rho\mu},\tilde{\Delta}_{\nu}\hat{f}_V\}_{\rm c}\Big),
\end{eqnarray}
and
\begin{eqnarray}
	S^{\mu\nu}_{(n)}\tilde{\Delta}_{\mu}\tilde{\Delta}_{\nu}\hat{f}_V=\frac{S^{\mu\nu}_{(n)}}{2}
	\Big(i[F_{\mu\nu},\hat{f}_V]_{\rm c}+\frac{1}{2}\{(D_{[\mu}F_{\beta\nu]}),\partial^{\beta}_{p}\hat{f}_V\}_{\rm c}+\frac{1}{4}[F_{\alpha[\mu}F_{\beta\nu]},\partial^{\alpha}_{p}\partial^{\beta}_{p}\hat{f}_V]_{\rm c}\Big).
\end{eqnarray}
On the other hand, for $\delta'(p^2)$ related terms, we acquire
\begin{eqnarray}\nonumber
	\delta'(p^2)S^{\mu\nu}_{(n)}p^{\rho}\{F_{\rho\mu},\tilde{\Delta}_{\nu}\hat{f}_V\}&=&-\frac{\delta'(p^2)p_{\nu}}{2}\{\tilde{F}^{\mu\nu},\tilde{\Delta}_{\mu}\hat{f}_V\}_{\rm c}
	+\frac{\delta'(p^2)p_{\mu}n_{\nu}}{2p\cdot n}\{\tilde{F}^{\mu\nu},p\cdot\tilde{\Delta}\hat{f}_V\}_{\rm c}
	\\
	&&
	-\frac{\delta(p^2)}{2p\cdot n}\{\tilde{F}^{\nu\beta}n_{\beta},\tilde{\Delta}_{\nu}\hat{f}_V\}_{\rm c}
\end{eqnarray}
by using the Schouten identity and $p^2\delta'(p^2)=-\delta(p^2)$. Also, one finds
\begin{eqnarray}
	\frac{\delta'(p^2)}{2} 
	p_{\nu}\tilde{\Delta}_{\mu}\{\tilde{F}^{\mu\nu},\hat{f}_{V}\}_{\rm c}=\frac{\delta'(p^2)}{2}p_{\nu}\Big(\{\tilde{F}^{\mu\nu},\tilde{\Delta}_{\mu}\hat{f}_{V}\}_{\rm c}
	+\frac{1}{2}[[F_{\mu\rho},\tilde{F}^{\mu\nu}],\partial^{\rho}_{p}\hat{f}_V]_{\rm c}\Big) 
\end{eqnarray}
by using
\begin{eqnarray}
	[A,\{B,C\}]=\{B,[A,C]\}+\{[A,B],C\},\quad
	\{A,\{B,C\}\}=\{B,\{A,C\}\}+[[A,B],C].
\end{eqnarray}
Finally, the last term is given by
\begin{eqnarray}\nonumber
	&&\frac{1}{4}\big(\partial_{p}^{\rho}p_{\nu}\delta'(p^2)\big)\{F_{\rho\mu},\{\tilde{F}^{\mu\nu},\hat{f}_{V}\}\}_{\rm c}
	\\\nonumber
	&&=\frac{1}{4}\big(\delta'(p^2)\{F_{\nu\mu},\{\tilde{F}^{\mu\nu},\hat{f}_{V}\}\}_{\rm c}+2p_{\nu}p^{\rho}\delta''(p^2)\{F_{\rho\mu},\{\tilde{F}^{\mu\nu},\hat{f}_{V}\}\}_{\rm c}\big)
	\\
	&&=\frac{1}{2}\delta''(p^2)p_{\nu}p^{\rho}[[\tilde{F}^{\mu\nu},F_{\rho\mu}],\hat{f}_V]_{\rm c},
\end{eqnarray}
where we have used
\begin{eqnarray}
	2p_{\nu}p^{\rho}\{F_{\rho\mu},\{\tilde{F}^{\mu\nu},\hat{f}_{V}\}\}_{\rm c}
	=-\frac{p^2}{2}\{F_{\mu\nu}\{\tilde{F}^{\mu\nu},\hat{f}_V\}\}_{\rm c}+2p_{\nu}p^{\rho}[[\tilde{F}^{\mu\nu},F_{\rho\mu}],\hat{f}_V]_{\rm c}
\end{eqnarray}
obtained from the Schouten identity and $p^2\delta''(p^2)=-2\delta'(p^2)$. Combining all terms together, We eventually derive a free-streaming effective AKE for massless quarks coupled with background color fields
\begin{eqnarray}\nonumber
	0&=&\delta(p^2)\bigg[p\cdot\tilde{\Delta}\hat{f}_A+\hbar(\partial_{\mu}S^{\mu\nu}_{(n)})\tilde{\Delta}_{\nu}\hat{f}_V+\frac{\hbar}{2p\cdot n}S^{\mu\nu}_{(n)}\{n^{\rho}F_{\mu\rho},\tilde{\Delta}_{\nu}\hat{f}_V\}_{\rm c}
	+\frac{\hbar S^{\mu\nu}_{(n)}}{2}
	\Big(i[F_{\mu\nu},\hat{f}_V]_{\rm c}
	\\\nonumber
	&&+\frac{1}{2}\{(D_{[\mu}F_{\beta\nu]}),\partial^{\beta}_{p}\hat{f}_V\}_{\rm c}+\frac{1}{4}[F_{\alpha[\mu}F_{\beta\nu]},\partial^{\alpha}_{p}\partial^{\beta}_{p}\hat{f}_V]_{\rm c}\Big)\bigg]
	+\frac{\hbar\delta'(p^2)p_{\mu}n_{\nu}}{2p\cdot n}\{\tilde{F}^{\mu\nu},p\cdot\tilde{\Delta}\hat{f}_V\}_{\rm c}
	\\
	&&+\frac{\hbar\delta'(p^2)}{4}p_{\nu}
	[[F_{\mu\rho},\tilde{F}^{\mu\nu}],\partial^{\rho}_{p}\hat{f}_V]_{\rm c}+\frac{\hbar\delta''(p^2)}{2}p_{\nu}p^{\rho}[[F_{\mu\rho},\tilde{F}^{\mu\nu}],\hat{f}_V]_{\rm c}.
\end{eqnarray}

\subsection{Spin diffusion and source terms}
Implementing the color decomposition $\hat{f}_{V/A}=\hat{f}_{V/A}^{\rm s}I+\hat{f}_{V/A}^at^a$ and taking $p\cdot\tilde{\Delta}\hat{f}_V=0$ from the off-shell SKE, the color-singlet and color-octet components of the AKE are given by 
\begin{eqnarray}\nonumber\label{AKE_singlet_massless}
	0&=&\delta(p^2)\bigg[p^{\mu}\mathcal{K}_{\rm s\mu}[\hat{f}_A]+\hbar(\partial_{\mu}S^{\mu\nu}_{(n)})\mathcal{K}_{\rm s\nu}[\hat{f}_V]
	+\frac{\hbar S^{\mu\nu}_{(n)}}{2p\cdot n}\bar{C}_{2}n^{\rho}F^{a}_{\mu\rho}\big(2\partial_{\nu}f_{V}^{a}+d^{abc}F^{b}_{\lambda\nu}\partial_{p}^{\lambda}\hat{f}_V^c\big)
	\\
	&&+\hbar S^{\mu\nu}_{(n)}
	\Big(\bar{C}_{2}\big((\partial_{\mu}F^{a}_{\beta\nu})-f^{abc}A^{b}_{\mu}F^{c}_{\beta\nu}\big)\partial_{p}^{\beta}\hat{f}_{V}^{a}\bigg],
\end{eqnarray}
and
\begin{eqnarray}\nonumber\label{AKE_octet_massless}
	0&=&\delta(p^2)\bigg[p^{\mu}\mathcal{K}^a_{\rm o \mu}[\hat{f}_A]+\hbar(\partial_{\mu}S^{\mu\nu}_{(n)})\mathcal{K}^a_{\rm o\mu}[\hat{f}_V]
	+\frac{\hbar}{2p\cdot n}S^{\mu\nu}_{(n)}n^{\rho}\big(2F^{a}_{\mu\rho}\mathcal{K}_{\rm s\nu}[\hat{f}_V]
	+d^{abc}F^{b}_{\mu\rho}\mathcal{K}^c_{\rm o\nu}[\hat{f}_V]\big)
	\\\nonumber
	&&+\hbar S^{\mu\nu}_{(n)}\big((\partial_{\mu}F^a_{\beta\nu})-f^{abc}A^b_{\mu}F^c_{\beta\nu}\big)\partial^{\beta}_{p}\hat{f}^{\rm s}_V
	\\\nonumber
	&&-\frac{\hbar S^{\mu\nu}_{(n)}}{2}
	\Big(f^{abc}F^b_{\mu\nu}
	-d^{abc}\big((\partial_{\mu}F^b_{\beta\nu})-f^{bef}A_{\nu}^eF^f_{\beta\nu}\big)\partial^{\beta}_{p}+\frac{f^{abc}f^{bef}}{4}F^e_{\alpha\mu}F^f_{\beta\nu}\partial^{\alpha}_{p}\partial^{\beta}_{p}\Big)\hat{f}^c_V\bigg]
	\\
	&&-\frac{\hbar}{4}p_{\nu}f^{abc}f^{bef}F^{e}_{\mu\rho}\tilde{F}^{f\mu\nu}\Big(\delta'(p^2-m^2)\partial^{\rho}_{p}+2p^{\rho}\delta''(p^2-m^2)\Big)\hat{f}_V^c,
\end{eqnarray}
where
\begin{eqnarray}\nonumber
	\mathcal{K}_{\rm s\mu}[O]&\equiv& \partial_{\mu} O^{s}+\bar{C}_{2}F^{a}_{\nu\mu}\partial_{p}^{\nu}O^{a},
	\\
	\mathcal{K}^a_{\rm o\mu}[O]&\equiv&\partial_{\mu} O^{a}-f^{bca}A^{b}_{\mu}O^{c}+F^a_{\nu\mu}\partial^{\nu}_{p}O^{\rm s}+\frac{d^{bca}}{2}F^b_{\nu\mu}\partial_{p}^{\nu}O^c.
\end{eqnarray}
Given $\mathcal{O}^a\sim\mathcal{O}(g)$, Eq.~(\ref{AKE_octet_massless}) leads to
\begin{align}
	0\approx p^{\mu}\big(\partial_{\mu} \hat{f}_A^a-f^{bca}A^{b}_{\mu}\hat{f}^{c}_A+F^a_{\nu\mu}\partial^{\nu}_{p}\hat{f}_V^{\rm s}\big)+\hbar(\partial_{\mu}S^{\mu\nu}_{(n)})\partial_{\nu}\hat{f}^a_V+\hbar S^{\mu\nu}_{(n)}\Big(\frac{n^{\rho}F^{a}_{\mu\rho}}{p\cdot n}\partial_{\nu}+(\partial_{\mu}F^{a}_{\beta\nu})\partial_{p}^{\beta}\Big)\hat{f}^{s}_{V}
\end{align}
and thus 
\begin{eqnarray}\nonumber
	\hat{f}_A^a(p,X)&=&-\int^{ab,p}_{k,X'}\Big[p^{\mu}F^b_{\nu\mu}(X')\partial^{\nu}_{p}+\hbar S^{\mu\nu}_{(n)}\Big(\frac{n^{\rho}F^{b}_{\mu\rho}(X')}{p\cdot n}\partial_{X'\nu}+\big(\partial_{X'\mu}F^{a}_{\beta\nu}(X')\big)\partial_{p}^{\beta}\Big)\Big]\hat{f}^{s}_{V}(p,X')
	\\
	&&+\int^{ab,p}_{k,X'}\hbar(\partial_{X'\mu}S^{\mu\nu}_{(n)})\partial_{X'\nu}\int^{bc,p}_{k',X''}\Big[p^{\alpha}F^c_{\beta\alpha}(X'')\partial^{\beta}_{X''p}\hat{f}^{\rm s}_V(p,X'')\Big]
\end{eqnarray}
up to $\mathcal{O}(g)$, where we have employed Eq.~(\ref{fV_octet_sol}) to rewrite $\hat{f}^a_{V}$ in terms of $\hat{f}^{s}_{V}$. For simplicity, we may work with a constant frame vector. 
Then Eq.~(\ref{AKE_singlet_massless}) becomes 
\begin{eqnarray}
	0&\approx&\delta(p^2)\bigg[p\cdot\partial \hat{f}^{\rm s}_A+\bar{C}_2p^{\mu}F^a_{\nu\mu}\partial^{\nu}_{p}\hat{f}^a_A
	+\frac{\hbar S^{\mu\nu}_{(n)}}{p\cdot n}\bar{C}_{2}n^{\rho}F^{a}_{\mu\rho}\partial_{\nu}\hat{f}_{V}^{a}
	+\hbar S^{\mu\nu}_{(n)}
	\bar{C}_{2}(\partial_{\mu}F^{a}_{\rho\nu})\partial_{p}^{\rho}\hat{f}_V^a\bigg]
\end{eqnarray} 
up to $\mathcal{O}(g^2)$. By replacing $\hat{f}^a_{V/A}$ with $\hat{f}^{\rm s}_{V/A}$ and taking the ensemble averages, we eventually derive
\begin{eqnarray}\label{CKE_signlet_simplify}
	0=\delta(p^2)\Big(p\cdot\partial f^{\rm s}_{A}(p,X)-\partial_{p}^{\kappa}\mathscr{D}_{\kappa}[f^{\rm s}_A]
	+\hbar\partial_{p}^{\sigma}\mathscr{B}_{\sigma}[f^{\rm s}_{V}]
	+\hbar S^{\mu\nu}_{(n)}\mathscr{C}_{\mu\nu}[f^{\rm s}_{V}]\Big),
\end{eqnarray}
where
\begin{align}
\mathscr{B}_{\sigma}[O]=\bar{C}_2p^{\rho}S^{\mu\nu}_{(n)}\int^p_{k,X'}\bigg[\big(\partial_{X'\mu}\langle F^a_{\sigma\rho}(X)F^a_{\nu\lambda}(X')\rangle\big) \partial_{p}^{\lambda}
-\frac{n^{\lambda}}{p\cdot n}\langle F^a_{\sigma\rho}(X)F^a_{\nu\lambda}(X')\rangle\partial_{X'\mu}
\bigg]O(p,X')
\end{align}
and
\begin{eqnarray}
	\mathscr{C}_{\mu\nu}[O]=\bar{C}_2p^{\rho}\int^p_{k,X'}\bigg[\frac{n^{\lambda}}{p\cdot n}\partial_{X'\mu}\langle F^a_{\nu\lambda}(X)F^a_{\sigma\rho}(X')\rangle
	+\partial_{X\mu}\langle F^a_{\nu\lambda}(X)F^a_{\sigma\rho}(X')\rangle \partial^{\lambda}_{p}\bigg]\partial^{\sigma}_{p}O(p,X').
\end{eqnarray} 
Here we have utilized
\begin{eqnarray}\nonumber
\bar{C}_2p^{\mu}\langle F^a_{\nu\mu}\partial^{\nu}_{p}\hat{f}^a_A\rangle&=&\hbar\bar{C}_2p^{\rho}\partial_{p}^{\sigma}\bigg\{S^{\mu\nu}_{(n)}\int^p_{k,X'}\bigg[\big(\partial_{X'\mu}\langle F^a_{\sigma\rho}(X)F^a_{\nu\lambda}(X')\rangle\big) \partial^{\lambda}_{p}
\\
&&-\frac{n^{\lambda}}{p\cdot n}\langle F^a_{\sigma\rho}(X)F^a_{\nu\lambda}(X')\rangle\partial_{X'\mu}\bigg]
f^{\rm s}_{V}(p,X')\bigg\},
\end{eqnarray}
\begin{eqnarray}\label{mag_term_AKE}
\frac{\hbar S^{\mu\nu}_{(n)}}{p\cdot n}\bar{C}_{2}n^{\rho}\langle F^{a}_{\mu\rho}\partial_{\nu}\hat{f}_{V}^{a}\rangle=
\frac{\hbar S^{\mu\nu}_{(n)}}{p\cdot n}\bar{C}_{2}n^{\lambda}\langle F^a_{\nu\lambda}(X)\partial_{X\mu}\int^p_{k,X'}F^a_{\sigma\rho}(X')\rangle\partial^{\sigma}_{p}f^{\rm s}_{V}(p,X'),
\end{eqnarray}
\begin{eqnarray}
\hbar S^{\mu\nu}_{(n)}
\bar{C}_{2}\langle (\partial_{\mu} F^{a}_{\rho\nu})\partial_{p}^{\rho}\hat{f}_V^a\rangle=\hbar \bar{C}_2S^{\mu\nu}_{(n)}\partial^{\lambda}_{p}\int^p_{k,X'}p^{\rho}\partial_{X\mu}\langle F^a_{\nu\lambda}(X)F^a_{\sigma\rho}(X')\rangle \partial_{p}^{\sigma}f^{\rm s}_{V}(p,X'),
\end{eqnarray}
and have assumed the relation
\begin{eqnarray}\label{dX_rel}
\partial_{X\mu}\int^p_{k,X'}G(X')=-i\int^p_{k,X'}k_{\mu}G(X')
=\int^p_{k,X'}\partial_{X'\mu}G(X')
\end{eqnarray}
given $e^{ik\cdot X'}G(X')\big|_{X'^{\mu}=\pm\infty}=0$ for an arbitrary function $G(X')$ is applicable for Eq.~(\ref{mag_term_AKE}). 

Although the $\hbar$ correction, corresponding to the source term for dynamical spin polarization, in the AKE for massless quarks takes a slightly different form compared to that for massive fermions, they both stem from spacetime variations of color fields. On the other hand, the color-singlet axial-vector component of the Wigner function reads
\begin{eqnarray}\nonumber\label{eq:Asmu_massless_exp}
	\langle \mathcal{A}^{\rm s\mu}\rangle&=&2\pi\bigg\{\delta(p^2)\bigg[p^{\mu}f^{\rm s}_{A}+\hbar S^{\mu\nu}_{(n)}\big(\partial_{\nu}f_{V}^{\rm s}+\bar{C}_{2}\langle F^{a}_{\rho\nu}\partial_{p}^{\rho}\hat{f}_V^a\rangle\big)
	-\frac{\hbar\bar{C}_2}{2}\langle\tilde{F}^{a\mu\nu}\partial_{p\nu}\hat{f}^a_{V}\rangle
	\bigg]
	\\
	&&+\frac{\hbar\bar{C}_2}{2}\partial_{p\nu}\big(\delta(p^2)\langle\tilde{F}^{a\mu\nu}\hat{f}^a_{V}\rangle\big)\bigg\},
\end{eqnarray}
which yields the color singlet of the on-shell axial charge current density in phase space, 
	\begin{eqnarray}\label{eq:Asmu_massless}
		\mathcal{A}^{\rm s\mu}(\bm p,X)\equiv \int \frac{dp_0}{2\pi}\langle\mathcal{A}^{\rm s\mu}\rangle
		=\frac{1}{2|\bm p|}\big(p^{\mu}f_A^{\rm s}+\hbar\bar{C}_2\tilde{\mathcal{A}}^{\mu}_{Q}\big)_{p_0=|\bm p|},
	\end{eqnarray}
where $\tilde{\mathcal{A}}^{\mu}_Q=\tilde{\mathcal{A}}^{\mu}_{Q_1}+\tilde{\mathcal{A}}^{\mu}_{Q2}+\tilde{\mathcal{A}}^{\mu}_{Q3}$ with  $\tilde{\mathcal{A}}^{\mu}_{Q1}=\mathcal{A}^{\mu}_{Q1}|_{m=0}$,  $\tilde{\mathcal{A}}^{\mu}_{Q2}=\mathcal{A}^{\mu}_{Q2}|_{m=0}$, and 
\begin{eqnarray}
\tilde{\mathcal{A}}^{\mu}_{Q3}(\bm p,X)
&=&-\bigg[S^{\mu\nu}_{(n)}\partial_{p}^{\rho}\int^p_{k,X'}p^{\beta}\langle F^{a}_{\rho\nu}(X)F_{\alpha\beta}(X')\rangle\partial^{\alpha}_{p}f_V^{\rm s}(p,X')\bigg]_{p_0=|\bm p|},
\end{eqnarray} 
which also contain the source terms coming from $\hbar$ corrections for spin polarization.

\subsection{Spin polarization, axial charge currents, and axial Ward identity}
We may now also investigate the spin polarization, axial charge current, and axial Ward identity of massless fermions for $f^{\rm s}_{V}$ near equilibrium with negligible gradient corrections in hydrodynamics. In such a case, it is more convenient to choose $n^{\mu}=u^{\mu}$ and work in the fluid rest frame \footnote{Similar to finding the local-equilibrium distribution functions in CKT \cite{Hidaka:2017auj}, such a frame choice is made and the distribution function in an arbitrary frame can be derived from the modified frame transformation.}. We re-emphasize that the Wigner function is however independent of the choice of a frame. Moreover, we may consider the color-field correlators only depending on $\delta X_0$ with the temporal direction now defined in the fluid rest frame. Notably, since $S^{\mu\nu}_{(n)}\partial_{\delta X\nu}\Phi(\delta X_0)=0$ for $n^{\mu}=u^{\mu}\approx(1,\bm 0)$, it turns out that the source term in Eq.~(\ref{CKE_signlet_simplify}) vanishes in our particular setup. In such a case, $f_A^{s}$ should simply diffuse to zero in equilibrium given no vortical corrections. When $f^{\rm s}_{V}$ is in equilibrium and $f^{\rm s}_A=0$, Eq.~(\ref{eq:Asmu_massless_exp}) reduces to
\begin{eqnarray}\label{Asmu_massless}
\hat{\mathcal{A}}^{\rm s\mu}(\bm p,X)=\hbar\bar{C}_2\big(\tilde{\mathcal{A}}^{\mu}_{Q1}(\bm p,X)+\tilde{\mathcal{A}}^{\mu}_{Q2}(\bm p,X)+\tilde{\mathcal{A}}^{\mu}_{Q3}(\bm p,X)\big),
\end{eqnarray}
where $\tilde{\mathcal{A}}^{\mu}_{Q1}(\bm p,X)+\tilde{\mathcal{A}}^{\mu}_{Q2}(\bm p,X)$ is nothing but Eq.~(\ref{AQ_massive}) with $m=0$, and
\begin{eqnarray}
	\tilde{\mathcal{A}}^{\mu}_{Q3}(\bm p,X)
	=\frac{\pi^{1/2}\tau_c}{2p_0}S^{\mu\nu}_{(n)}\langle E^{a}_{\nu}E^{a}_{\beta}\rangle \hat{p}^{\beta}\big(1-p_{0}\partial_{p0}\big)\partial_{p0}f_{\rm eq}(p_0)\Big|_{p_0=|\bm p|}
\end{eqnarray} 
by similarly assuming
$\Phi(\delta X_0)=e^{-\delta X_0^2/\tau_{c}^2}$ and working in the fluid-rest frame. When dropping the subleading correlation function of two color electric fields, the non-dynamical source term for spin polarization of massless quarks matches the one of massive quarks by simply taking $m=0$.

One can analogously evaluate the axial charge current for finite $\tau_c$, which reads
\begin{eqnarray}\nonumber
	J^{\mu}_5(X)&=&-2\hbar \int\frac{d^4p}{(2\pi)^3}\frac{\delta(p^2)}{2}\int^{p}_{k,X'}
	\langle B^{a\rho}(X)E^a_{\rho}(X')\rangle \beta u^{\mu}f_{\rm eq}(p\cdot u)(1-f_{\rm eq}(p\cdot u))
	\\
	&=&2\hbar\int\frac{d^4p}{(2\pi)^3}\frac{\delta(p^2)}{4p\cdot u}\sqrt{\pi}\tau_c
	\langle E^a\cdot B^a\rangle u^{\mu}\partial_{p\cdot u}f_{\rm eq}(p\cdot u)
\end{eqnarray} 
and agrees with the result for massive fermions by taking $m=0$ and hence $\partial\cdot J_5=0$. In the limit for $\tau_c\rightarrow \infty$, following the previous calculations for massive fermions and taking $\langle B^{a\rho}(X)E^a_{\rho}(X')\rangle=\langle B^{a}\cdot E^a\rangle$, it turns out that
\begin{eqnarray}\label{divJ5_anomaly}
	\partial_{\mu} J^{\mu}_5(X)
	=-\frac{\hbar}{4\pi^2}\langle B^{a}\cdot E^a\rangle,
\end{eqnarray}
which corresponds to the non-Abelian axial anomaly as expected. 

\section{Summary and discussions}\label{sec:summary}
Here we briefly summarize the primary findings. We have derived the AKE containing a classical spin diffusion term and a source term from quantum corrections that could possibly trigger spin polarization for massive quarks (with mass greater than the gradient scales) as shown in Eq.~(\ref{AKE_signlet_simplify}). A similar equation for massless quarks is also derived in Eq.~(\ref{CKE_signlet_simplify}). 
On the other hand, there also exist explicit quantum corrections in Wigner functions, which severe as the additional source terms to generate spin polarization shown in Eqs.~(\ref{eq:Asmu_massive}) and (\ref{eq:Asmu_massless}) for massive and massless quarks, respectively. Given postulated color-field correlators based on spacetime translational invariance and spatial homogeneity, the parity-odd correlators of color fields, depending on their time difference can dominantly generate nonzero spin polarization as shown in Eq.~(\ref{Amu_equil}) for massive fermions near thermal equilibrium. Such a contribution further results in a nonzero axial charge current or more precisely the axial charge density in the fluid rest frame. This constant axial charge current also gives the vanishing axial Ward identity with finite correlation time of the color-field correlators. However, in the constant-field limit with infinite correlation time, the nonzero axial Ward identity associated with the pseudo-scalar condensate at finite temperature is acquired as shown in Eq.~(\ref{divJ5_condensate}). For massless fermions, a particular frame choice leads to the vanishing dynamical source term in AKE, while there exists a similar non-dynamical source term for spin polarization in the Wigner function. Note that the choice of a frame is simply for the technical reason, which does not affect the physical observables in the end. The axial Ward identity reduces to the expected axial anomaly as manifested by Eq.~(\ref{divJ5_anomaly}) in the constant-field limit as anticipated. In order to make a smooth connection between the results of the massive and massless quarks, it is necessary to further work out the AKT of massive quarks in a proper frame for arbitrary mass such as $n^{\mu}=u^{\mu}$ for $f^s_V$ near equilibrium. 

In phenomenology, as proposed in Ref.~\cite{Muller:2021hpe}, such spin polarization of quarks engendered by color-field correlators could potentially affect the spin alignment of vector mesons observed in heavy ion collisions \cite{ALICE:2019aid,Singha:2020qns} as an unsettled question in theory (see Refs~.\cite{Sheng:2019kmk,Sheng:2020ghv,Xia:2020tyd,Goncalves:2021ziy} for other theoretical explanations). In order to make direct comparisons with experimental data, it is necessary to incorporate the color-field correlators from real-time simulations with prescribed initial conditions. According to the approximations adopted in Ref.~\cite{Muller:2021hpe} and this work, the parity-odd correlator between a chromo-magnetic field and a chromo-electric field is essential to generate spin polarization for massive quarks. However, due to the event-by-event fluctuations of the sign for such a correlator, the anomalous spin polarization by turbulent color fields may contribute to spin alignment of vector mesons, coming from the product of spin polarization for a quark and for an anti-quark, without affecting the spin polarization of Lambda hyperons. Further studies on simulations or more practical estimation of color-field correlators and the generalization of present QKT with source terms to arbitrary quark mass will be needed.

\acknowledgments
The author would like to thank B. M\"uller for insightful comments on the preliminary version of this work and K. Hattori for fruitful discussions. This work was supported by Ministry of Science and Technology, Taiwan under Grant No. MOST 110-2112-M-001-070-MY3.

\appendix
\section{Details of integration}\label{app:integration}
In this appendix, we will consider the relevant integral of an arbitrary function $\tilde{G}(p,X,X')=G(p,X,\delta X)$. For brevity, we will simply denote $G(p,X,\delta X)=G(X,\delta X)$, while the $p$ dependence has to be taken into account in the end. We now consider the integral
\begin{eqnarray}
I_{k,X'}\equiv\int d^4k\int\frac{d^4X'}{(2\pi)^4}\frac{e^{ik\cdot(X'-X)}}{p\cdot k}\tilde{G}(X,X')
=\int d^4k\int\frac{d^4\delta X}{(2\pi)^4}\frac{e^{-ik\cdot \delta X}}{p\cdot k}G(X,\delta X),
\end{eqnarray}
where $\delta X\equiv X-X'$. Assigning $p_{\mu}=(p_0,0,0,p_z)$, it is found
\begin{eqnarray}\nonumber
I_{k,X'}&=&\int\frac{d^4\delta X}{(2\pi)^2}\int dk_0dk_z\frac{e^{-ik_0\delta X_0+ik_z\delta X_z}}{p_0k_0-p_zk_z}\delta^2(\delta X_{\rm T})G(X,\delta X)
\\\nonumber
&=&\int\frac{d^4\delta X}{(2\pi)^2}\int dk_zd\tilde{k}_0\frac{e^{-i\tilde{k}_0\delta X_0-i\frac{p_zk_z}{p_0}\delta X_0+ik_z\delta X_z}}{p_0\tilde{k}_0}\delta^2(\delta X_{\rm T})G(X,\delta X)
\\
&=&-i\pi\int\frac{d^4\delta X}{(2\pi)}\delta\left(\frac{p_z}{p_0}\delta X_0-\delta X_z\right)\frac{{\rm sgn}(\delta X_0)}{p_0}\delta^2(\delta X_{\rm T})G(X,\delta X),
\end{eqnarray}
where $\delta X_{\rm T}$ are the transverse components of $\delta X_{\mu}$ with respect to $p_{\mu}$, $\tilde{k}_0=k_0-p_zk_z/p_0$, and we have utilized the Fourier transform 
\begin{eqnarray}
\int_{-\infty}^{\infty} dxe^{-ikx}/x=-i\pi{\rm sgn}(k).
 \end{eqnarray}
Integrating over $\delta X_{\rm T}$ and $\delta X_{z}$, one finally obtains 
\begin{eqnarray}\label{Intkxp}
I_{k,X'}=-\frac{i}{2p_0}\int d\delta X_0\Big[{\rm sgn}(\delta X_0)G(X,\delta X)\Big]_{\delta X_{T}=0,\delta X_z=p_z\delta X_0/p_0}.
\end{eqnarray}
It turns out that $I_{k,X'}$ is non-vanishing only when $G(X,\delta X)$ is an odd function of $\delta X_0$ or of $\delta X_z$. 

On the contrary, we shall consider the integral
\begin{eqnarray}
	I^{+}_{k,X'}\equiv
	\lim_{\epsilon\rightarrow 0}\int d^4k\int\frac{d^4\delta X}{(2\pi)^4}\frac{e^{-ik\cdot \delta X}}{p\cdot k+i\epsilon}G(X,\delta X),
\end{eqnarray}
where $\epsilon>0$. By employing the decomposition,
\begin{eqnarray}
\frac{1}{p\cdot k+i\epsilon}=\frac{p\cdot k}{(p\cdot k)^2+\epsilon^2}-\frac{i\epsilon}{(p\cdot k)^2+\epsilon^2},
\end{eqnarray} 
we may also decompose $I^{+}_{k,X'}$ into
\begin{eqnarray}
I^{+}_{k,X'}=I^{+(1)}_{k,X'}+I^{+(2)}_{k,X'},
\end{eqnarray}
where 
\begin{eqnarray}
I^{+(1)}_{k,X'}=\lim_{\epsilon\rightarrow 0}\int d^4k\int\frac{d^4\delta X}{(2\pi)^4}\frac{p\cdot ke^{-ik\cdot \delta X}}{(p\cdot k)^2+\epsilon^2}G(X,\delta X)
\end{eqnarray}
and
\begin{eqnarray}
	I^{+(2)}_{k,X'}=\lim_{\epsilon\rightarrow 0}\int d^4k\int\frac{d^4\delta X}{(2\pi)^4}\frac{-i\epsilon e^{-ik\cdot \delta X}}{(p\cdot k)^2+\epsilon^2}G(X,\delta X).
\end{eqnarray}
Using 
\begin{eqnarray}
	\lim_{\epsilon\rightarrow 0}\frac{\epsilon}{x^2+\epsilon^2}=\pi\delta(x),
\end{eqnarray}
one immediately finds
\begin{eqnarray}\label{I2+int}
	I^{+(2)}_{k,X'}=-i\pi\int d^4k\int\frac{d^4\delta X}{(2\pi)^4}\delta(p\cdot k) e^{-ik\cdot \delta X}G(X,\delta X).
\end{eqnarray}
To evaluate $I^{+(1)}_{k,X'}$, we assign $p_{\mu}=(p_0,0,0,p_z)$ and hence
\begin{eqnarray}
	I^{+(1)}_{k,X'}=\lim_{\epsilon\rightarrow 0}\int dk_0dk_z\int\frac{d\delta X_0d\delta X_z}{(2\pi)^2}\frac{(p_0k_0-p_zk_z)e^{-ik_0\delta X_0+ik_z\delta X_z}}{(p_0k_0-p_zk_z)^2+\epsilon^2}G(X,\delta X)|_{\delta X_{x,y}=0},
\end{eqnarray}
where we have integrated over $k_{x,y}$ and then $\delta X_{x,y}$. We then first take the limit of the integrand,
\begin{eqnarray}
\lim_{\epsilon\rightarrow 0}\frac{(p_0k_0-p_zk_z)e^{-ik_0\delta X_0+ik_z\delta X_z}}{(p_0k_0-p_zk_z)^2+\epsilon^2}=\frac{e^{-ik_0\delta X_0+ik_z\delta X_z}}{(p_0k_0-p_zk_z)}
\end{eqnarray}
and subsequently integrate over $k_0$ by using
\begin{eqnarray}
	\int_{-\infty}^{\infty} dke^{-ikx}/(k+a)=-i\pi{\rm sgn}(x)e^{ika}.
\end{eqnarray}
It is found
\begin{eqnarray}\nonumber
	I^{+(1)}_{k,X'}&=&-i\pi\int dk_z\int\frac{d\delta X_0d\delta X_z}{(2\pi)^2}\frac{{\rm sgn}(\delta X_0)}{p_0}e^{ik_z(\delta X_z-\delta X_0p_z/p_0)}G(X,\delta X)|_{\delta X_{x,y}=0}
	\\\nonumber
	&=&-i\pi\int\frac{d\delta X_0d\delta X_z}{(2\pi)}\frac{{\rm sgn}(\delta X_0)}{p_0}\delta\big(\delta X_z-\delta X_0p_z/p_0\big)G(X,\delta X)|_{\delta X_{x,y}=0}
	\\
	&=&-\frac{i}{2p_0}\int d\delta X_0\Big[{\rm sgn}(\delta X_0)G(X,\delta X)\Big]_{\delta X_{x,y}=0,\delta X_z=p_z\delta X_z/p_0}.
\end{eqnarray}
which agrees with Eq.~(\ref{Intkxp}). Similarly, from Eq.~(\ref{I2+int}), one can also obtain 
\begin{eqnarray}
I^{+(2)}_{k,X'}&=&-\frac{i}{2p_0}\int d\delta X_0G(X,\delta X)|_{\delta X_{x,y}=0,\delta X_z=p_z\delta X_0/p_0},
\end{eqnarray}
and hence 
\begin{eqnarray}
	I^{+}_{k,X'}&=&-\frac{i}{2p_0}\int d\delta X_0\Big(1+{\rm sgn}(\delta X_0)\Big)G(X,\delta X)|_{\delta X_{x,y}=0,\delta X_z=p_z\delta X_0/p_0},
\end{eqnarray}
Eventually, the principle value part of $1/(p\cdot k+i\epsilon)$ should still contribute to the integral $I^{+}_{k,X'}$. Nevertheless, when $G(X,\delta X)$ is an even function of $\delta X_{0,z}$, only $I^{+(2)}_{k,X'}$ yields a non-vanishing result. The conclusion is then consistent with the finding in Ref.~\cite{Asakawa:2006jn}.  

Recall that $G(X,\delta X)=G(p,X,\delta X)$. Given
\begin{eqnarray}
I^+_{k,X'}=-i\int^{p}_{k,X'}G(X,\delta X)=-\frac{i}{2p_0}\int d\delta X_0\Big(1+{\rm sgn}(\delta X_0)\Big)G(X,\delta X)|_{\delta X_c},
\end{eqnarray}	
where $\delta X_c\equiv\{\delta X_{x,y}=0,\delta X_z=p_z\delta X_0/p_0\}$,
it is found
\begin{eqnarray}
	\partial_{p\kappa}I^{+}_{k,X'}&=&-\frac{\delta^{0}_{\kappa}}{p_0}I^+_{k,X'}-\frac{i}{2p_0}\int d\delta X_0\Big(1+{\rm sgn}(\delta X_0)\Big)\partial_{p\kappa}\big(G(X,\delta X)|_{\delta X_c}\big).
\end{eqnarray}	
Using the chain rule,
\begin{eqnarray}
\partial_{p\kappa}\big(G(X,\delta X)|_{\delta X_c}\big)=\big(\partial_{p\kappa}G(X,\delta X)\big)_{\delta X_c}+\big(\partial_{\delta X_z}G(X,\delta X)\big)_{\delta X_c}\delta X_0\partial_{p\kappa}(p_z/p_0).
\end{eqnarray}
one finds another useful relation,
\begin{eqnarray}\nonumber\label{eq:DkintG}
\partial_{p\kappa}\int^{p}_{k,X'}G(X,\delta X)&=&-\frac{\delta^{0}_{\kappa}}{p_0}\int^{p}_{k,X'}G(X,\delta X)
+\int^{p}_{k,X'}\partial_{p\kappa}G(X,\delta X)
\\
&&+\int^{p}_{k,X'}\delta X_0\frac{\partial G(X,\delta X)}{\partial \delta X^{\mu}_{\perp}}\partial_{p\kappa}\left(\frac{p_{\perp}^{\mu}}{p_0}\right)
\end{eqnarray}
for the off-shell integral. When imposing the on-shell condition first, an analogous relation reads 
\begin{eqnarray}\nonumber\label{eq:DkintG_onshell}
	\partial_{p\kappa}\Big(\int^{p}_{k,X'}G(X,\delta X)\Big)_{p_0=\epsilon_{\bm p}}&=&-\frac{p_{\perp\kappa}}{\epsilon_{\bm p}^2}\int^{\bm p}_{k,X'}G(X,\delta X)_{p_0=\epsilon_{\bm p}}
	+\int^{\bm p}_{k,X'}\partial_{p_{\perp}\kappa}G(X,\delta X)_{p_0=\epsilon_{\bm p}}
	\\
	&&+\int^{\bm p}_{k,X'}\delta X_0\left(\frac{\partial G(X,\delta X)}{\partial \delta X^{\mu}_{\perp}}\right)_{p_0=\epsilon_{\bm p}}\partial_{p_{\perp}\kappa}\left(\frac{p_{\perp}^{\mu}}{\epsilon_{\bm p}}\right),
\end{eqnarray}
where $\int^{\bm p}_{k,X'}=\int^{p}_{k,X'}\big|_{p_0=\epsilon_{\bm p}}$.

\bibliography{QKT_quarks_arXiv_v2.bbl}
\end{document}